\documentclass[aps,pra,twocolumn,amsmath,amssymb,superscriptaddress,floatfix]{revtex4-1}

\usepackage{graphicx}
\usepackage{color}
\usepackage{makecell}
\usepackage{xspace}

\usepackage{hyperref}

\newcommand{\tket}{\ensuremath{\mathsf{t}|\mathsf{ket}\rangle}\xspace}

\begin{document}


\title{Dynamical mean field theory algorithm and experiment on quantum computers}


\author{I. Rungger}
\email[]{ivan.rungger@npl.co.uk}
\affiliation{National Physical Laboratory, Teddington, TW11 0LW, United Kingdom}

\author{N. Fitzpatrick}
\affiliation{Cambridge Quantum Computing Ltd, 9a Bridge Street, Cambridge, United Kingdom}

\author{H. Chen}
\affiliation{Rahko Ltd., Finsbury Park, N4 3JP, United Kingdom}
\affiliation{Dept. of Computer Science, University College London, Gower Street, London, WC1E 6BT, United Kingdom}

\author{C. H. Alderete}
\affiliation{Joint Quantum Institute, Department of Physics, University of Maryland, College Park, MD 20742, USA}
\affiliation{Instituto Nacional de Astrof\'{i}sica, \'{O}ptica y Electr\'{o}nica, Calle Luis Enrique Erro No. 1, Sta. Ma. Tonantzintla, Pue. CP 72840, Mexico  }

\author{H. Apel}
\affiliation{Cambridge Quantum Computing Ltd, 9a Bridge Street, Cambridge, United Kingdom}
\affiliation{Dept. of Physics and Astronomy, University College London, Gower Street, London, WC1E 6BT, United Kingdom}

\author{A. Cowtan}
\affiliation{Cambridge Quantum Computing Ltd, 9a Bridge Street, Cambridge, United Kingdom}

\author{A. Patterson}
\affiliation{Dept. of Physics and Astronomy, University College London, Gower Street, London, WC1E 6BT, United Kingdom}

\author{D. Mu\~{n}oz Ramo}
\affiliation{Cambridge Quantum Computing Ltd, 9a Bridge Street, Cambridge, United Kingdom}

\author{Y. Zhu}
\affiliation{Joint Quantum Institute, Department of Physics, University of Maryland, College Park, MD 20742, USA}

\author{N. H. Nguyen}
\affiliation{Joint Quantum Institute, Department of Physics, University of Maryland, College Park, MD 20742, USA}

\author{E. Grant}
\affiliation{Rahko Ltd., Finsbury Park, N4 3JP, United Kingdom}
\affiliation{Dept. of Computer Science, University College London, Gower Street, London, WC1E 6BT, United Kingdom}

\author{S. Chretien}
\affiliation{National Physical Laboratory, Teddington, TW11 0LW, United Kingdom}

\author{L. Wossnig}
\affiliation{Rahko Ltd., Finsbury Park, N4 3JP, United Kingdom}
\affiliation{Dept. of Computer Science, University College London, Gower Street, London, WC1E 6BT, United Kingdom}

\author{N. M. Linke}
\affiliation{Joint Quantum Institute, Department of Physics, University of Maryland, College Park, MD 20742, USA}

\author{R. Duncan}
\affiliation{Cambridge Quantum Computing Ltd, 9a Bridge Street, Cambridge, United Kingdom}
\affiliation{Department of Computer and Information Sciences, University of Strathclyde, 26 Richmond Street, Glasgow, United Kingdom}


\begin{abstract}
The developments of quantum computing algorithms and experiments for atomic scale simulations have largely focused on quantum chemistry for molecules, while their application in condensed matter systems is scarcely explored.
Here we present a quantum algorithm to perform dynamical mean field theory (DMFT) calculations for condensed matter systems on currently available quantum computers, and demonstrate it on two quantum hardware platforms. DMFT is required to properly describe the large class of materials with strongly correlated electrons. The computationally challenging part arises from solving the  effective problem of an interacting impurity coupled to a bath, which scales exponentially with system size on conventional computers. An exponential speedup is expected on quantum computers, but the algorithms proposed so far are based on real time evolution of the wavefunction, which requires high-depth circuits and hence very low noise levels in the quantum hardware. Here we propose an alternative approach, which uses the variational quantum eigensolver (VQE) method for ground and excited states to obtain the needed quantities as part of an exact diagonalization impurity solver. We present the algorithm for a two site DMFT system, which we benchmark using simulations on conventional computers as well as experiments on superconducting and trapped ion qubits, demonstrating that this method is suitable for running DMFT calculations on currently available quantum hardware.
\end{abstract}


\maketitle


\section{Introduction}

Computers have an integral role in the design process of new materials and chemicals, predicting and explaining the behavior of the systems in question~\cite{macalino2015role,aaqvist1994new,parr1980density,Jain2016}.
Such simulations are mostly based on density functional theory (DFT) due to its low computational cost and simplicity of use~\cite{macalino2015role}. DFT allows for the computation of important properties, such as atomic structures, chemical reaction rates and the electronic structure of solids.
However,  in many instances  DFT fails to correctly predict the observed behavior~\cite{cohen2008insights}, for example Mott insulator transitions \cite{imada1998metal} or the binding of oxygen to haemoglobin~\cite{weber2013importance,weber2014renormalization}.
The reason is that DFT treats the quantum mechanical interactions between electrons within an effective mean field approximation, which becomes invalid when the electrons are strongly correlated with each other. 
A number of corrections have been developed to overcome the limitations of DFT \cite{Mardirossian2017}, such as hybrid functionals \cite{HSE06}, self-interaction corrections~\cite{Pertsova}, or the GW approximation \cite{Caruso}. 
For solid state systems, where local electron-electron correlations are strong, the dynamical mean field theory (DMFT) is the state-of-the-art correction to DFT~\cite{me.vo.89,ge.ko.96,ko.vo.04}.
In DMFT one separates out the strongly interacting local
orbitals as an effective impurity from the remaining part of the system, which is treated as an effective bath together with the orbitals of neighboring atoms.
Example cases where DMFT overcomes the failures of DFT include Mott insulators \cite{imada1998metal} and superconducting systems \cite{bednorz1986possible}, molecules on surfaces exhibiting Kondo behavior \cite{Droghetti.17}, and even extend to biochemical systems \cite{kovaleva2008versatility} and metalloproteins such as haemoglobin~\cite{weber2014renormalization}.
Metalloproteins play an important role within the pharmaceutical sector, such as the therapeutic application of hemocyanins~\cite{gesheva2014anti}, a very promising class of anti-cancer therapeutics.

However, the high computational cost required for accurate solutions limits DMFT to small systems on the currently available conventional computers.
Quantum computers can in principle solve correlated electronic structure problems in polynomial time~\cite{lloyd1996universal,aspuru2005simulated,lanyon2010towards}
using algorithms such as phase estimation~\cite{kitaev9511026quantum}.
While large-scale quantum computers are still out of reach, small, so called noisy intermediate scale quantum computers~\cite{preskill2018quantum} have recently become available, and have sparked a growing interest in demonstrating applications on existing devices.
In particular, quantum-classical hybrid algorithms, such as the variational quantum eigensolver (VQE)~\cite{peruzzo},
have been shown to enable simulations for small molecules and simplified model Hamiltonians \cite{Kandala2019,whitfield2011simulation}, perform learning tasks~\cite{benedetti2019parameterized,benedetti2019adversarial,mitarai2018quantum,chen2018universal} or even algebraic operations~\cite{xu2019variational}.
However, due to the noise levels and limited size of current quantum computers simulations of large molecules or materials systems are still out of reach. Embedding methods such as DMFT may overcome this problem, since here the computationally demanding part of the calculation is performed only on a smaller subsystem.
A recent proposal demonstrates that such methods can also be formulated using quantum algorithms~\cite{Bauer2016,wecker2015solving}.
In these approaches the computationally intensive part of the DMFT calculation is executed on a quantum device, while the remaining part is done classically.
This hybrid approach has been simulated on a conventional computer for a small prototype system of 2 sites~\cite{Kreula2016} within the framework of 2-site DMFT~\cite{Potthoff2001}.

To date no hybrid quantum-classical algorithm that solves the DMFT formalism has been implemented on an actual quantum device, because prior methods have been unable to deal with the high levels of noise in these machines.
Here we expand upon prior work by introducing a VQE-based quantum-classical hybrid algorithm, which allows us to compute the 2-site DMFT system on a quantum device.
The quantum computer solves the effective quantum impurity problem, which in the DMFT loop is self-consistently determined via a feedback between the quantum and classical computation.
We use VQE to implement an exact diagonalization solver \cite{Caffarel1994,Qimiao1994,Liebsch2012}, 
and demonstrate that our algorithm can tackle the electronic structure problem of  correlated materials using quantum devices available today.
This method can be scaled to larger system sizes as the available number of qubits increases,
where its efficiency relies on the scalability of the underlying VQE, which is an active area of research \cite{wecker2014gate,huggins2019efficient}.
With that, we believe our method can allow many open questions in quantum materials to be resolved once an intermediate scale quantum computer with around 100 qubits becomes available.

\section{Quantum algorithm}
\subsection{Dynamical mean field theory}
\label{sec:dmft}
Within DMFT the extended lattice model is mapped to an effective Anderson impurity problem with Hamiltonian operator $\hat{H}$, where the interacting region is coupled to an infinitely extended bath \cite{me.vo.89,ge.ko.96,ko.vo.04}. Here we use the exact diagonalization (ED) approximation as the impurity solver, where a finite number of effective sites is used to represent the bath \cite{Caffarel1994,Qimiao1994,Liebsch2012}. To map this ED Hamiltonian to a quantum computer we perform a Jordan-Wigner transform \cite{PhysRevLett.63.322}. As a matter of notation, we label quantities associated with the $N_\mathrm{imp}$ impurity spin orbitals with Greek subscripts $\alpha,\beta,\gamma,\delta$. The spin orbital index is a collective index that includes both orbital and spin degrees of freedom, so that for example $\alpha=1$ represents the spin orbital $(1,\uparrow)$, where the first integer indicates the site index and the arrow the spin direction, $\alpha=2$ represents $(1,\downarrow)$, $\alpha=3$ represents $(2,\uparrow)$, and so on, until the index $\alpha=2 N_\mathrm{imp}$ represents $(N_\mathrm{imp},\downarrow)$.
The spin orbitals associated with the $N_\mathrm{b}$ bath orbitals are labeled with $i,j$ subscripts, and the ones associated with any of the $N_\mathrm{imp}+N_\mathrm{b}$ orbitals with $n$ and $m$ subscripts. 
Within ED the impurity Hamiltonian is usually written as \cite{Liebsch2012,Liebsch2007,Bauer2016} 
\begin{align}
\hat{H}&=\hat{H}_\mathrm{imp}+\hat{H}_\mathrm{bath}+\hat{H}_\mathrm{mix},\label{eq:H}\\
\hat{H}_\mathrm{imp}&=\sum_{\alpha}\left(\epsilon_{\alpha}-\mu\right)\hat{\sigma}_\alpha^- \hat{\sigma}_\alpha^+ +\sum_{\alpha\beta\gamma\delta}U_{\alpha\beta\gamma\delta}\hat{\sigma}_\alpha^- \hat{\sigma}_\beta^- \hat{\sigma}_\gamma^+ \hat{\sigma}_\delta^+,\\
\hat{H}_\mathrm{mix}&=\sum_{\alpha i}\left(V_{\alpha i} \hat{\sigma}_{\alpha}^- \hat{\sigma}_i^+ +V_{\alpha i}^* \hat{\sigma}_i^- \hat{\sigma}_{\alpha}^+\right),\\
\hat{H}_\mathrm{bath}&=\sum_{i}\epsilon_i \hat{\sigma}_i^- \hat{\sigma}_i^+,
\end{align}
where $\mu$ is the chemical potential, $\epsilon_{\alpha}$ are the internal onsite energies of the impurity, $U_{\alpha\beta\delta\gamma}$ are the electron interaction energies, $V_{\alpha i}$ are the hopping matrix elements between the impurity and the bath, and the $\epsilon_i$ denote the onsite energies of the bath orbitals. As part of the standard Jordan-Wigner transform here we have introduced a modified form of the Pauli ladder operators that takes into account the fermionic nature of the electrons
\begin{align}
\hat{\sigma}_{\alpha}^\pm&=\left(\prod_{\beta=1}^{\alpha-1}\hat{\sigma}_\beta^z\right) \frac{1}{2} \left(\hat{\sigma}_\alpha^x \pm i \hat{\sigma}_\alpha^y\right),\\
\hat{\sigma}_{i}^\pm&=\left(\prod_{\beta=1}^{N_\mathrm{imp}}\hat{\sigma}_\beta^z\right) \left(\prod_{j=1}^{i-1}\hat{\sigma}_j^z\right) \frac{1}{2} \left(\hat{\sigma}_i^x \pm i \hat{\sigma}_i^y\right).
\end{align}
With this definition $\hat{\sigma}_{\alpha(i)}^-$ [$\hat{\sigma}_{\alpha(i)}^+$] creates [destroys] an electron on spin orbital $\alpha(i)$. The electron number operator on spin orbital $\alpha(i)$ is then given by $\hat{n}_{\alpha(i)}=\hat{\sigma}_{\alpha(i)}^- \hat{\sigma}_{\alpha(i)}^+$, so that for $\left<\hat{\sigma}_{\alpha(i)}^z\right>=1$ [$\left<\hat{\sigma}_{\alpha(i)}^z\right>=-1$] spin orbital $\alpha(i)$ is empty [filled].
The total number of qubits to represent $\hat{H}$ is $2(N_\mathrm{imp}+N_\mathrm{b})$.

We can calculate all the energy eigenvalues of $\hat{H}$, $E_{N,n}$, and the corresponding eigenvectors, $\left.|\psi_{N,n}\right>$. Here the integer $N$ denotes the number of electrons of the state, and the integer $n$ goes from 0 to the number of eigenstates with $N$ electrons minus one, $M_N$. We order the states by increasing energy, so that $\left.|\psi_{N,0}\right>$ is the ground state for $N$ electrons, $\left.|\psi_{N,1}\right>$ is the first excited state and so on, until the highest excited state for N electrons $\left.|\psi_{N,M_N}\right>$. We denote as $N_0$ the number of electrons of the overall ground state, and label the ground state (GS) as $\left.|\psi_0\right>=\left.|\psi_{N_0,0}\right>$, with $E_0=E_{N_0,0}$.

\begin{figure}
\centering\includegraphics[width=0.47\textwidth,clip=true]{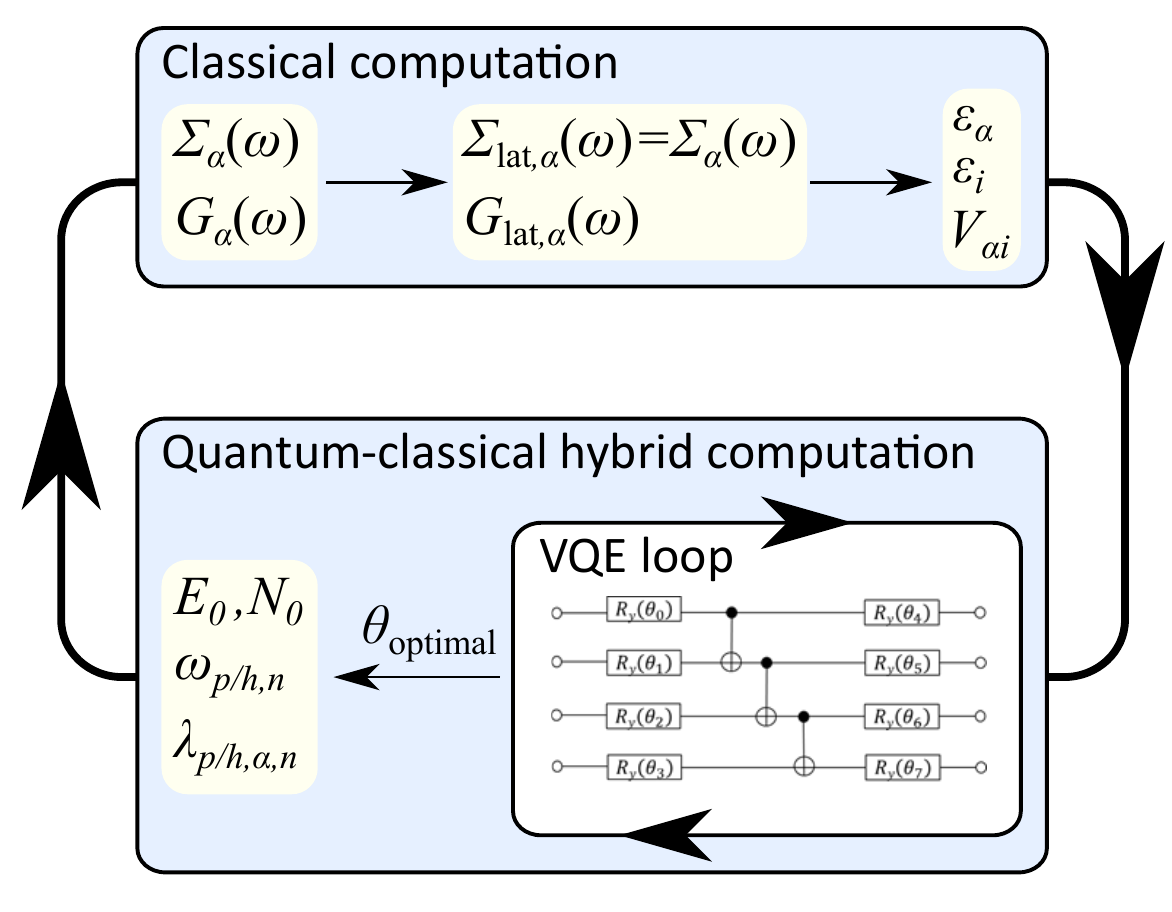}
\caption{Schematic of the ED DMFT loop: we start by choosing initial values of the impurity Hamiltonian parameters $\lbrace \epsilon_{\alpha}, \epsilon_{i},V_{\alpha i}\rbrace$, which we then use as input to the quantum computing part of the loop; here we use a VQE quantum circuit to obtain as output all the $\lbrace \omega_{\mathrm{p/h},n}, \lambda_{\mathrm{p/h},\alpha,n}\rbrace$. These are then passed as input to the classical computing part of the loop, where the GFs and self-energies are computed to obtain updated parameters $\lbrace  \epsilon_{\alpha}, \epsilon_{i}, V_{\alpha i}\rbrace$. These are then again passed to the quantum computing part of the loop, and the process is iterated until the $\lbrace \epsilon_{\alpha}, \epsilon_{i}, V_{\alpha i}\rbrace$ are equal within a specified tolerance between subsequent iterations. Once this DMFT self-consistency is achieved, one can use the obtained self-consistent impurity Hamiltonian parameters to calculate the electronic structure of the system.
}
\label{fig:DMFTSchematic}
\end{figure}

Within DMFT the central quantities are the local retarded Green's functions (GFs) of the original lattice model, $G_\mathrm{lat}(\omega)$, and of the impurity problem, $G(\omega)$. Here $\omega$ is the real or imaginary energy. The central condition of DMFT is that locally on the interacting site the impurity problem is equivalent to the lattice problem, which leads to the condition that $G_\mathrm{lat}(\omega)=G(\omega)$.   
This condition can only be exactly satisfied if an infinite number of bath sites is used. For a finite number of bath sites in ED one typically aims at effectively minimizing the difference between $G_\mathrm{lat}(\omega)$ and $G(\omega)$.  
In practice one starts from some initial guess for impurity Hamiltonian parameters $\epsilon_\alpha$, $\epsilon_i$ and $V_{\alpha i}$, and uses them to calculate G($\omega$). From these quantities and the original lattice Hamiltonian one obtains $G_\mathrm{lat}(\omega)$. At this stage then one updates $\epsilon_\alpha$, $\epsilon_i$ and $V_{\alpha i}$ to effectively aim to minimize the difference between $G(\omega)$ and $G_\mathrm{lat}(\omega)$. One then iterates this procedure until $\epsilon_\alpha$, $\epsilon_i$ and $V_{\alpha i}$ reach self-consistency. This process is denoted as the DMFT loop (Fig. \ref{fig:DMFTSchematic})\cite{Liebsch2012}.

We now assume that the local GF's off-diagonal terms are small enough that they can be neglected. Note that we use this assumption only to simplify the notation, and one can equivalently formulate the algorithm also for dense GFs.
We use the Lehman representation of the diagonal elements of the zero temperature impurity GF, $G_{\alpha}(\omega)$, which are generally given by
\begin{equation}
G_{\alpha}(\omega)=\sum_{n=0}^{M_{N_0-1}}  \frac{\lambda_{\mathrm{h},\alpha,n}}{\omega+i\delta-\omega_{\mathrm{h},n}} + \sum_{n=0}^{M_{N_0+1}}  \frac{\lambda_{\mathrm{p},\alpha,n}}{\omega+i\delta-\omega_{\mathrm{p},n}},
\label{eq:g}
\end{equation}
where the first summation goes over all states with one electron removed from the ground state (hole states), while the second summation goes over all states with one electron added to it (particle states)\cite{Liebsch2012}.
Here $\delta$ is an infinitesimally small positive number, and
\begin{align}
\omega_{\mathrm{h},n}&=E_{0}-E_{N_0-1,n},\label{eq:omegah}\\
\lambda_{\mathrm{h},\alpha,n}&=\left|\left<\psi_{N_0-1,n}\left|\left(\prod_{\beta=1}^{\alpha-1}\hat{\sigma}_\beta^z\right)\hat{\sigma}_\alpha^x\right|\psi_0\right>\right|^2,\label{eq:alphah}\\
\omega_{\mathrm{p},n}&=E_{N_0+1,n}-E_{0},\\
\lambda_{\mathrm{p},\alpha,n}&=\left|\left<\psi_{N_0+1,n}\left|\left(\prod_{\beta=1}^{\alpha-1}\hat{\sigma}_\beta^z\right)\hat{\sigma}_\alpha^x\right|\psi_0\right>\right|^2.\label{eq:alphap}
\end{align}
Note that in Eqs. (\ref{eq:alphah}) and (\ref{eq:alphap}) we use the fact that the number of electrons of the states in the brackets differs by one (see Appendix \ref{sec:AppendixLambda}). The matrix elements $\lambda_{\mathrm{p/h},\alpha,n}$ have values between zero and one, and satisfy the sum rule $\sum_n (\lambda_{\mathrm{p},\alpha,n}+ \lambda_{\mathrm{h},\alpha,n})=1$. For increasing $N_\mathrm{imp}+N_\mathrm{b}$ the total number of excited states $M_{N_0\pm1}$ becomes exponentially large. However, in practical calculations one typically needs only the low energy excitations up to a specified energy cutoff, which are needed to represent $G_\alpha(\omega)$ in the desired energy range around 0.

We can equivalently write $G_{\alpha}(\omega)$ as
\begin{equation}
G_\alpha(\omega)=\frac{1}{\omega+i\delta-\epsilon_\alpha+\mu-\Delta_\alpha(\omega)-\Sigma_\alpha(\omega)},
\end{equation}
where the so-called hybridization function is given by
\begin{equation}
\Delta_\alpha(\omega)=\sum_i\frac{\left|V_{\alpha i}\right|^2}{\omega+i\delta-\epsilon_{i}},
\label{eq:delta}
\end{equation}
and we have introduced the many-body self-energy, $\Sigma_\alpha(\omega)$, which includes all the modifications to the non-interacting Green's function, $G_{0,\alpha}(\omega) =\left[\omega+i\delta-\epsilon_\alpha+\mu-\Delta_\alpha(\omega)\right]^{-1}$, induced by the interaction term proportional to $U$ in the Hamiltonian.
It can be written in general form as 
$\Sigma_\alpha(\omega)=G_{0,\alpha}^{-1}(\omega)-G_\alpha^{-1}(\omega)$.
In our quantum algorithm we determine $G_\alpha(\omega)$ with calculations on a quantum computer, which means that we evaluate the quantities in Eqs. (\ref{eq:omegah}-\ref{eq:alphap}) on a quantum device. The remaining computations of the DMFT calculation are performed on a classical computer (Fig. \ref{fig:DMFTSchematic}).

\subsection{Quantum Circuits}
\label{sec:qc}
In principle $\omega_{\mathrm{p/h},n}$ and $\lambda_{\mathrm{p/h},\alpha,n}$ can be obtained by performing a Fourier transform of the real-time GF \cite{Kreula2016,Bauer2016}. However, the error induced by the finite time steps used to evaluate the real-time GF grows for larger interaction strengths, and therefore the method requires very small time steps. On the currently available quantum computers this leads to noise levels in the results that do not allow for accurate computations of these quantities.
A recent proposal to mitigate the errors in the Trotter evolution is to use effectively averaged integrated quantities \cite{Keen2019}.
Our proposed method is to calculate $\omega_{\mathrm{p/h},n}$ and $\lambda_{\mathrm{p/h},\alpha,n}$ via total energy calculations based on a variational quantum eigensolver (VQE), which is generally more resilient to noise \cite{peruzzo, PhysRevLett.119.180509, Kandala2019, PhysRevX.7.021050}. This approach has also been suggested in a recent article for general Green's function based calculations \cite{SuguruEndo}. In the following we only present the method to calculate $\omega_{\mathrm{p},n}$ and $\lambda_{\mathrm{p},\alpha,n}$, since the procedure to obtain $\omega_{\mathrm{h},n}$ and $\lambda_{\mathrm{h},\alpha,n}$ is analogous.

We represent a general eigenstate of $\hat{H}$ on the quantum computer by
\begin{equation}
\left.|\psi_{N,n}\right>=\hat{U}_{N,n}\left.|0\right>,
\label{eq:psiU}
\end{equation}
where $\hat{U}_{N,n}$ is a unitary operator, and the $\left.|0\right>$ state represents the state with $\left<\hat\sigma^z_m\right>=1$ for all $m$. We typically use a hardware efficient ansatz to apply $\hat{U}_{N,n}$ to the state $\left.|0\right>$ in our quantum circuit \cite{HardwareEffAns}.
Note that for the common case where the parameters in $\hat{H}$ are real, all the expansion coefficients of the $\left.|\psi_{N,n}\right>$ can be made real as well, so that $\hat{U}_{N,n}$ can be restricted to real operators. We therefore typically do not need to include any $R_x$ rotations in our ansatz, since these would only be required if the coefficients have a complex component, and instead use only $R_y$ rotations. 

We first prepare the state $\left.|\psi_0\right>$ by varying the parameters of the ansatz quantum circuit to minimize the total energy and obtain $E_0$ and with it $N_0$. 
To then calculate the spectrum of the $N_0+1$ and $N_0-1$ electron states we use a modified Hamiltonian,
\begin{equation}
\hat{\tilde{H}}=\hat{H}+\beta \left(\hat{N}-N_\mathrm{target}\right)^2.
\label{eq:HN}
\end{equation}
 Here $\hat{N}=\sum_{\alpha} \hat{n}_\alpha + \sum_{i} \hat{n}_i $ is the total number operator.
In $\hat{\tilde{H}}$ a penalty term proportional to a real parameter $\beta$ is added to enforce the target electron number $N_\mathrm{target}$ \cite{ryabinkin_penalties, qiskit_webpage}, in our case $N_\mathrm{target}=N_0\pm1$.
To calculate the ground state and excited states of $\hat{\tilde{H}}$ one can then use a number of methods \cite{SSVQExcitedState, Higgott2019variationalquantum,Jones_ExcitedStates2019}. Here we use the algorithm proposed in Ref. [\onlinecite{Higgott2019variationalquantum}], which finds the excited states by penalizing the overlap of the higher energy eigenstates with the previously found lower energy eigenstates.

With this method Eqs. (\ref{eq:omegah}-\ref{eq:alphap}) can be evaluated. To calculate $\lambda_{\mathrm{p},\alpha,n}$ without increasing the number of qubits we combine Eqs. (\ref{eq:alphap}) and (\ref{eq:psiU}) to
\begin{equation}
\lambda_{\mathrm{p},\alpha,n}=\left|\left<0\left|\hat{U}^\dagger_{N_0+1,n}\left(\prod_{\beta=1}^{\alpha-1}\hat{\sigma}_\beta^z\right)\hat{\sigma}_\alpha^x\hat{U}_{N_0,0}\right|0\right>\right|^2.
\label{eq:lambdaUsigmaU}
\end{equation}
We therefore obtain $\lambda_{\mathrm{p},\alpha,n}$ on a quantum device by evaluating the quantum circuit corresponding to $\hat{U}^\dagger_{N_0+1,\alpha}\left(\prod_{\beta=1}^{\alpha-1}\hat{\sigma}_\beta^z\right)\hat{\sigma}_\alpha^x \hat{U}_{N_0,0}$ and measuring the probability of finding the $\left.|0 \right> $ state at the end.

\subsection{Regularization\label{sec:reg}}

With the quantities in Eqs. (\ref{eq:omegah}-\ref{eq:alphap}) calculated on a quantum computer one can in principle perform the full DMFT loop in Fig. \ref{fig:DMFTSchematic}. However, the small deviations of the calculated quantities from the exact values lead to the presence of unphysical poles in $\Sigma(\omega)$, and hence usually the DMFT loop does not converge. These unphysical divergences are found at the energies where $G_{0,\alpha}(\omega)$ is 0, since $\Sigma_\alpha(\omega)=G_{0,\alpha}^{-1}(\omega)-G_\alpha^{-1}(\omega)$. The exact interacting $G_\alpha^{-1}(\omega)$ has divergences at the same energies, which exactly cancel out the divergences due to $G_{0,\alpha}^{-1}(\omega)$. For the approximate $G_\alpha(\omega)$ calculated with the quantum circuit the cancellation is only partial. We therefore need to perform a regularization of the calculated quantities, in which these are modified in order to restore the exact cancellation of the divergences of $G_{0,\alpha}(\omega)$. The set of energies, at which $G_{0,\alpha}(\omega)=0$, corresponds to the one where $\Delta_\alpha(\omega)$ in Eq. (\ref{eq:delta}) has a pole, and hence is equal to the set $\left\{\epsilon_i\right\}$. Cancellation of the divergences in $\Sigma_\alpha(\omega)$ therefore requires that $G_\alpha(\epsilon_i)=0$ and $\left.dG_\alpha(\omega)/d\omega=dG_{0,\alpha}(\omega)/d\omega\right|_{\omega=\epsilon_i}$ for all $\epsilon_i$. Together with Eq. (\ref{eq:g}) we then obtain the sum rules
\begin{align}
\sum_{n=0}^{M_{N_0-1}}  \frac{\lambda_{\mathrm{h},\alpha,n}}{\epsilon_i-\omega_{\mathrm{h},n}} + \sum_{n=0}^{M_{N_0+1}}  \frac{\lambda_{\mathrm{p},\alpha,n}}{\epsilon_i-\omega_{\mathrm{p},n}}=0.\label{eq:reg1}\\
\sum_{n=0}^{M_{N_0-1}}  \frac{\lambda_{\mathrm{h},\alpha,n}}{\left(\epsilon_i-\omega_{\mathrm{h},n}\right)^2} + \sum_{n=0}^{M_{N_0+1}}  \frac{\lambda_{\mathrm{p},\alpha,n}}{\left(\epsilon_i-\omega_{\mathrm{p},n}\right)^2}=\frac{1}{V_i^2}.\label{eq:reg2}
\end{align}
In principle one can therefore evaluate all quantities on a quantum computer, and then perform a constrained minimization procedure, where all $\lambda_{\mathrm{p/h},\alpha,n}$ and $\omega_{\mathrm{p/h},n}$ are modified by as little as possible to satisfy the sum rules above. This can be a challenging task for increasing system size. We therefore propose to perform the constrained minimization only on the $\lambda_{\mathrm{p/h},\alpha,n}$, while fixing the $\omega_{\mathrm{p/h},n}$. The reason is that in general we expect that the $\omega_{\mathrm{p/h},n}$ are calculated more accurately than the $\lambda_{\mathrm{p/h},\alpha,n}$, since the quantum circuit for the $\lambda_{\mathrm{p/h},\alpha,n}$ is longer. Since Eqs. (\ref{eq:reg1}) and (\ref{eq:reg2}) are linear in the $\lambda_{\mathrm{p/h},\alpha,n}$, the constrained minimization is straight forward even for larger system sizes \cite{NumericalOptimization2006}.
Overall the regularization is therefore scalable to large system sizes.

\section{2-site DMFT simulations and experiments}

\subsection{2-site DMFT model}
\label{sec:2sDMFTModel}

As a proof of concept we consider the 2-site DMFT system presented in Refs. [\onlinecite{Potthoff2001}] and [\onlinecite{Kreula2016}], which solves the single-band Hubbard model on the Bethe lattice with infinite connectivity using exact diagonalization of a two site impurity problem with one interacting and one bath site. Here we only present the parts required to perform the quantum computing DMFT calculations, for a detailed description of the model we refer to Ref. [\onlinecite{Potthoff2001}].
For this system $\hat{H}$ from Eq. (\ref{eq:H}) becomes
\begin{align}
\hat{H}&=\frac{U}{4}\hat{\sigma}_1^z \hat{\sigma}_3^z+\left(\frac{\mu}{2}-\frac{U}{4}\right)\left(\hat{\sigma}_1^z+ \hat{\sigma}_3^z\right)-\frac{\epsilon_2}{2}\left(\hat{\sigma}_2^z+ \hat{\sigma}_4^z\right)\nonumber\\
&+\frac{V}{2}\left(\hat{\sigma}_1^x \hat{\sigma}_2^x+\hat{\sigma}_1^y \hat{\sigma}_2^y+\hat{\sigma}_3^x \hat{\sigma}_4^x+\hat{\sigma}_3^y \hat{\sigma}_4^y\right).
\label{eq:H2s}
\end{align}
Note that here we use a modified mapping of indices between spin orbitals and qubits from that used in Eq. (\ref{eq:H}) in order to slightly reduce the number of resulting Pauli operators: qubit 1 (2) represents $\uparrow$ electrons on site 1 (2), while qubit 3 (4) represents a $\downarrow$ electron on site 1 (2).
We can then introduce the operator $\hat{S}_z$ for the total $z$ component of the spin as (in units of $\hbar/2$)
\begin{equation}
\hat{S}_z=\hat{n}_1 + \hat{n}_2 - \hat{n}_3 - \hat{n}_4,
\end{equation}
with expectation value $S_z=\langle\hat{S}_z\rangle$. Since $\hat{H}$ is equivalent for up- and down-spin electrons we have $G_3=G_1$ and $\Sigma_3=\Sigma_1$, so that we need to evaluate the Green's function and self-energy only for one spin (we evaluate $G=G_1$ and $\Sigma=\Sigma_1$).

In 2-site DMFT there are only two bath parameters that can be optimized in the DMFT self-consistent loop, $\epsilon_2$ and $V=V_{12}$. The conditions for DMFT self-consistency derived in Ref. [\onlinecite{Potthoff2001}] are
\begin{align}
n_\mathrm{imp}&=n_\mathrm{lat},\label{eq:DMFT1}\\
V^2&=z,\label{eq:DMFT2}
\end{align}
where we have implicitly set the unit of energy equal to the unscaled hopping of the Bethe lattice.
Here  $n_\mathrm{imp}=\int_{-\infty}^0\mathrm{DOS}_\mathrm{imp}(\omega)d\omega$ is the occupation of the impurity, with the impurity density of states (DOS) given by 
$\mathrm{DOS}_\mathrm{imp}(\omega)=-\frac{2}{\pi}\mathrm{Im}\left[G(\omega+i \delta)\right]$,
and $n_\mathrm{lat}=\int_{-\infty}^0\mathrm{DOS}_\mathrm{lat}(\omega)d\omega$ is the occupation of one Bethe lattice site in the periodic system, with the DOS of the Bethe lattice
$\mathrm{DOS}_\mathrm{lat}(\omega)=2 \rho_0\left[\omega +\mu-\Sigma(\omega)\right]$, and $\rho_0(x)=(1/2\pi)\sqrt{4-x^2}$.
In Eq. (\ref{eq:DMFT2}) We have introduced the quasi-particle weight $z$, also known as the wave-function re-normalization factor, defined as $z=1/\left(1-\left.\frac{d\mathrm{Re}[\Sigma(\omega)]}{d\omega}\right|_{\omega=0}\right)=1/\left(1-\frac{\mathrm{Im}\left[\Sigma(i\delta)\right]}{\delta}\right)$. 2-site DMFT self-consistency therefore requires updating $\epsilon_2$ and $V$ until Eqs. (\ref{eq:DMFT1}) and (\ref{eq:DMFT2}) are satisfied within a desired tolerance.

\subsection{Quantum computing implementation}
\label{sec:qci}

For our simulations in this section we consider the particle-hole (ph) symmetric case, where $\mu=U/2$ and $\epsilon_2=0$, so that the condition in Eq. (\ref{eq:DMFT1}) is automatically satisfied, and we only have to update $V$ until $\left|V-\sqrt{z}\right| < \eta$ to satisfy the condition in Eq. (\ref{eq:DMFT2}), with $\eta$ a small finite tolerance \cite{Kreula2016}.
Note that our scheme is generally applicable also away from ph symmetry, and we present an example for that in Appendix \ref{sec:AppendixNPHS}.
At each step of the iterative procedure we have a fixed pair of $U$ and $V$ as input to the quantum computation, which then provides as output the corresponding values for $\omega_{\mathrm{p/h},n}$ and $\lambda_{\mathrm{p/h},\alpha,n}$.
At ph symmetry we have $\omega_{\mathrm{h},n}=-\omega_{\mathrm{p},n}$ and $\lambda_{\mathrm{h},\alpha,n}=\lambda_{\mathrm{p},\alpha,n}$, so that we need to perform the computations only for the particle contributions.
To demonstrate the applicability of our DMFT method on current quantum hardware we therefore need to show that these quantities computed in an experiment deviate only little from the exact numerical results, which we do in the remaining part of the manuscript.
\begin{figure}[t]
\centering\includegraphics[width=0.47\textwidth,clip=true]{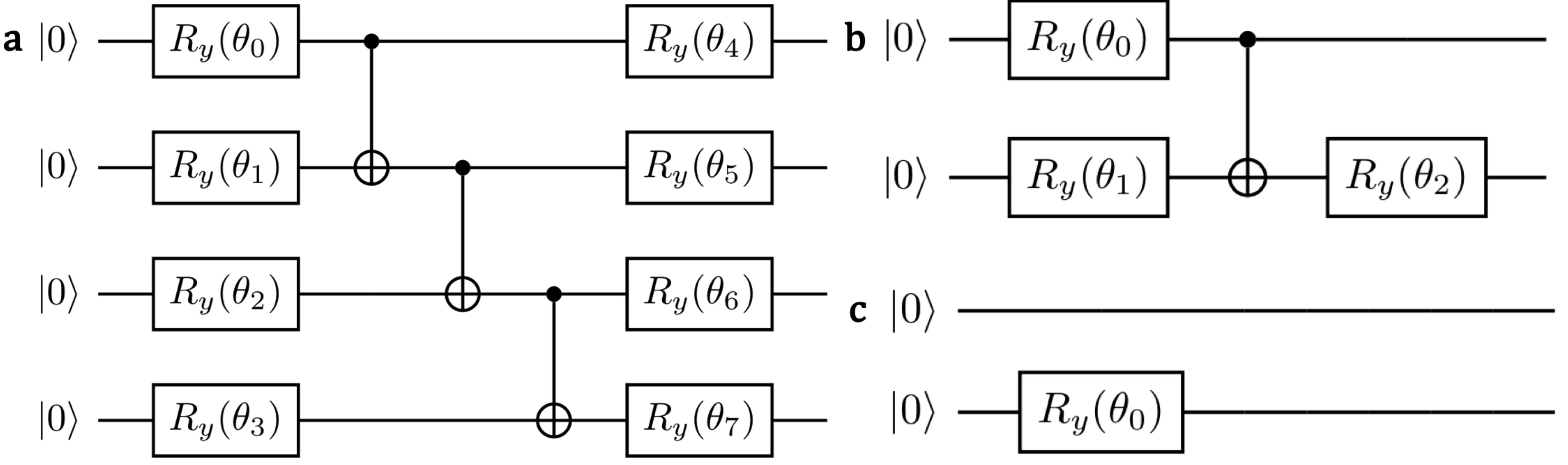}
\caption{a) Quantum circuit for the 4-qubit DMFT calculation based on the Penalty Term (PT) approach; b) and c) 2-qubit circuits used in the Circuit Reduction (CR) approach: b) circuit for the 2 electron $S_z=0$ ground state calculation; c) circuit for the 1 and 3 electron spectrum calculations. Since the expansion coefficients of our states are real, we only use $R_y$ rotations.}
\label{fig:QC}
\end{figure}

We independently use both the approach of circuit reduction (CR) and of the additional penalty term (PT) in the Hamiltonian (Eq.~(\ref{eq:HN})) to obtain energies at the required number of electrons (see Sec.~\ref{sec:qc}). Within the PT approach we use the full 4-qubit Hamiltonian in Eq.~(\ref{eq:H2s}) to calculate $E_0$, $N_0$ and $\hat{U}_{N_0,0}$, and then the states for $N_0+1$ electrons to obtain all quantities in Eqs.~(\ref{eq:omegah}-\ref{eq:alphap}). To represent the operators $\hat{U}_{N,n}$ on the quantum computer we use a circuit based on the hardware efficient approach for real Hamiltonians, with two layers of rotation gates and one layer of entangling gates, giving a total of 8 angles to be optimized (Fig. \ref{fig:QC}a).

For the CR approach we design a separate reduced circuit for each different $N$ and $S_z$, with a corresponding mapping to a reduced Hamiltonian (see Appendix \ref{sec:appendixCR}). In this case no penalty term is needed, and one can just run normal VQE on the individual circuits to determine the ground state, and subsequently calculate the excited states for the $N_0\pm1$ circuits. For 2-site DMFT the circuits can be reduced to 2 qubits for all $N$, which then also allows the calculation of the overlap terms with Eq. (\ref{eq:lambdaUsigmaU}) (see Appendix. \ref{sec:appendixCR}). For one and three electrons the resulting effective 2-qubit Hamiltonian has identical eigenvalues due to particle-hole symmetry ($E_{3,n}=E_{1,n}$), and is
$\hat{H}=\frac{U}{4}\hat{\sigma}_2^z+V \hat{\sigma}_2^x$ for one electron, and $\hat{H}=-\frac{U}{4}\hat{\sigma}_2^z+V \hat{\sigma}_2^x$ for three electrons. The state of the first qubit does not affect the Hamiltonian, reflecting the fact that there are always two degenerate states, one with $S_z=1$ and one with $S_z=-1$. The 2 qubit circuit required to obtain all the possible eigenvalues of these Hamiltonians is shown in Fig.~\ref{fig:QC}c, and consists of only one rotation; the degenerate eigenstate with opposite $S_z$ is then obtained by adding a $R_y(\pi)$ rotation on the first qubit. For two electrons there is one state with $S_z=2$, one with $S_z=-2$, and four with $S_z=0$, for which the Hamiltonian mapped to 2 qubits is
$\hat{H}=\frac{U}{4}\hat{\sigma}_1^z \hat{\sigma}_2^z+V\left(\hat{\sigma}_1^x+ \hat{\sigma}_2^x\right)$. The general 2-qubit ansatz for an eigenstate of this Hamiltonian is shown in Fig. \ref{fig:QC}b. Note that the CR approach is generally preferable over the PT approach, since it requires fewer angles as optimization parameters, and hence fewer gates and fewer qubits. The drawback is that it requires to find the projection of the Hamiltonian to a reduced set of qubits for the specified number of electrons and other conserved quantities, which can be a challenging task for a general Hamiltonian.

Since the Hamiltonian is spin-symmetric, for $N=3$ electrons we have 2 doubly degenerate independent energy eigenvalues ($E_{3,0}=E_{3,1}$ and $E_{3,2}=E_{3,3}$), where one has $S_z=1$ and one $S_z=-1$. For convenience we order the states in such a way that when they are degenerate, for $N=3$ the smaller $n$ refers to the $S_z=-1$ state ($S_z=1$ for $N=1$), and the larger $n$ to the $S_z=1$ state ($S_z=-1$ for $N=1$).  The poles in the GF (Eq. \ref{eq:g}) are then found at $\omega_{\mathrm{p},0}=\omega_{\mathrm{p},1}=E_{3,0}-E_0$ and $\omega_{\mathrm{p},2}=\omega_{\mathrm{p},3}=E_{3,2}-E_0$, as well as at $-\omega_{\mathrm{p},0}$ and $-\omega_{\mathrm{p},2}$ due to ph symmetry.
Furthermore, we have $\lambda_{\mathrm{p},1,1}=\lambda_{\mathrm{p},1,3}=0$, since these involve matrix elements between states with different spin.
Since $\sum_n (\lambda_{\mathrm{p},\alpha,n}+ \lambda_{\mathrm{h},\alpha,n})=1$, we then obtain $\lambda_{\mathrm{p},1,2}=1/2-\lambda_{\mathrm{p},1,0}$,
so that we have only one independent parameter, $\lambda=\lambda_{\mathrm{p},1,0}$. To solve the DMFT problem we therefore need to calculate the three energies $E_0$, $E_{3,0}$ and $E_{3,2}$, as well as $\lambda$ on the quantum computer.

\begin{table}[t]
\caption{Results of simulations on a classical computer without noise, comparing exact numerical results with those on a simulator for 4 qubits using the PT approach and for 2 qubits using the CR approach, for $U=4$ and $V=0.745356$. The ``$\infty$ Shots'' VQE runs use the information directly from the wave-function state vector rather than simulations of individual measurements; the obtained ``Optimal $\theta$'' angles are then used for a single step calculation with 5000 shots to indicate the statistical noise induced by a finite number of measurements. These optimal angles are then used also for the fixed-parameters runs on quantum hardware.
\label{tab:enesSimul}
}
\begin{tabular}{c|c|cc|cc}
\hline
         &  &        \multicolumn{2}{c|}{4 qubits}                         & \multicolumn{2}{c}{2 qubits}  \\
         & Exact        & \thead{VQE \\ ($\infty$ Shots)} & \thead{Optimal $\theta$ \\ (5000 Shots)}   &  \thead{VQE \\ ($\infty$ Shots)}   & \thead{Optimal $\theta$ \\ (5000 Shots)}    \\
\hline                                                                                      
 $E_0$    & -1.795     &  -1.795   & -1.804  &  -1.795  & -1.811  \\ 
 $E_{3,0}$& -1.247     &  -1.247   & -1.232  &  -1.247  & -1.279  \\
 $E_{3,2}$&  1.247     &   1.247   &  1.260  &   1.247  &  1.244  \\
$\lambda$ &  0.262     &   0.262   &  0.288  &   0.262  &  0.273  \\
\hline
\end{tabular}
\end{table}

\begin{figure}[t]
\centering\includegraphics[width=0.47\textwidth,clip=true]{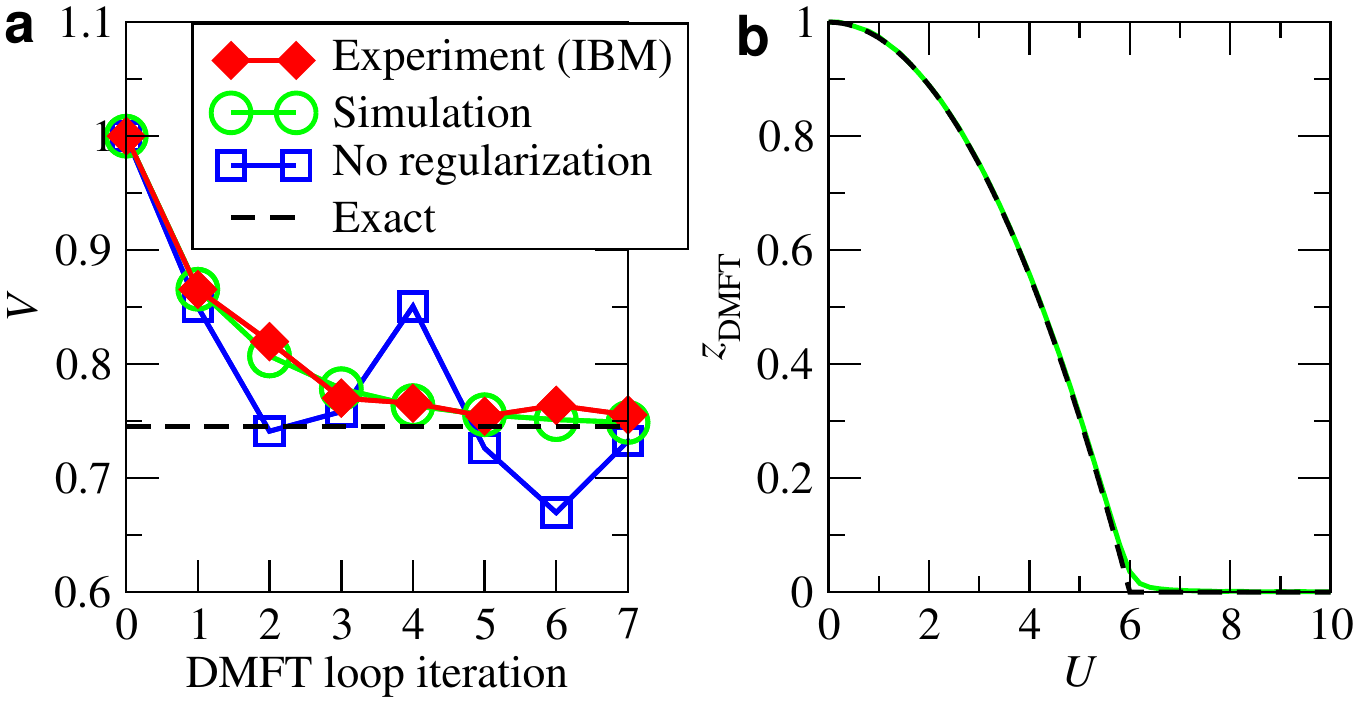}
\caption{a) Simulation and experiment on the IBM quantum computing hardware of the DMFT algorithm for the change of the bath parameter $V$ as function of DMFT loop iteration for a fixed value of $U=4$. Using the regularization procedure to calculate $\lambda$ (Eq. (\ref{eq:lambda})) the bath parameter $V$ converges quickly and smoothly to the correct value, and the experimental data follow rather closely the simulated results; without regularization no convergence is possible. b) Value of the quasi-particle weight $z$ at DMFT self-consistency obtained in simulations, where $z\approx V^2$. The exact numerical value agrees well with the numerical result obtained using the quantum simulation.
}
\label{fig:DMFTLoop}
\end{figure}
One important advantage of the chosen system is that it has exact analytical solutions for the ph-symmetric system (away from ph-symmetry we can still obtain essentially exact numerical solutions), which we can therefore use to benchmark the quality of the quantum computations. The analytic DMFT solution for $V$ is $V=\sqrt{z}=\sqrt{1-\left(\frac{U}{6}\right)^2}$ for $U< 6$, and $V=\sqrt{z}=0$ for $U\ge 6$, with $N_0=2$ and $S_z=0$ as GS. As first benchmarking test of the quantum algorithm we use a value of $U=4$ and the corresponding analytic DMFT solution $V=0.745356$, and compute $E_0$, $E_{3,0}$, $E_{3,2}$, and $\lambda$ on the quantum computer.
We first compute the GS using VQE with simulations of the algorithm on a classical computer, and for both the PT and the CR approaches we find that the GS is the $N_0=2$ and $S_z=0$, in agreement with the exact numerical result. The values of $E_0$ for the quantum simulation with effectively infinite statistical measurements are identical to the exact numerical results, and the same is true for $E_{3,0}, E_{3,2}$ and $\lambda$ (see Tab. \ref{tab:enesSimul}), showing that our ansatz circuits can fully represent the eigenstates. For this simulation we directly use the actual wave function of the system to evaluate the expectation values, which corresponds to the information obtained after an infinite amount of measurements. Any deviation from the exact results on a quantum device can therefore be attributed to the finite number of measurements (``shots'') and noise in the quantum computer.
We can then fix the angles $\theta$ in the circuit to the optimal values obtained with the VQE calculations for an effectively infinite number of shots, and perform a single step calculation with a finite number of shots at these ``Optimal $\theta$'' values.
For 5000 shots the discrepancies in the energy are within about 2\%, and for $\lambda$ they are also within about 10\%.

Next we perform the full DMFT loop simulation using the quantum circuit. Importantly, we need to regularize the calculated $\lambda$ to satisfy the sum rule constraints derived in Sec. \ref{sec:reg}, which in this case simplify to
\begin{equation}
\lambda=\frac{\omega_{\mathrm{p},0}^2\left(V^2-\omega_{\mathrm{p},2}^2\right)}{2V^2\left(\omega_{\mathrm{p},0}^2-\omega_{\mathrm{p},2}^2\right)}.
\label{eq:lambda}
\end{equation}
When applying regularization in the DMFT loop we therefore do not need to calculate $\lambda$ on the quantum computer, since it is obtained with this relation. In this case we only need to obtain $\omega_{\mathrm{p},0}$ and $\omega_{\mathrm{p},2}$ via total energy calculations on the quantum computer. We find that in this way the DMFT loop with the quantum algorithm converges well for all $U$, and is only limited by the statistical noise of the finite number of shots, while without regularization convergence cannot usually be achieved (Fig. \ref{fig:DMFTLoop}). This is due to the appearance of unphysical divergences in the self-energy when Eq. (\ref{eq:lambda}) is not exactly fulfilled,  which the regularization removes, as discussed in Sec. \ref{sec:reg}. Here we use 10000 shots to obtain the expectation values.
The calculated values of $z$ at DMFT self-consistency, $z_\mathrm{DMFT}$, agree well with the exact analytical ones for all values of $U$ (Fig. \ref{fig:DMFTLoop}b). The only minor difference is that the transition from finite to zero $z$ is somewhat smoothed out in the simulated results due to the fact that we only use a finite number of shots for the evaluation of expectation values.

\subsection{Experiments on quantum computers}

To demonstrate that the algorithm runs successfully on current quantum computers we first perform the calculation for the pair $U=4$ and $V=0.745356$ on two independent types of quantum computing hardware, the IBM superconducting qubit architecture and the ion trap quantum computer at the University of  Maryland. We then perform a full DMFT loop on the IBM quantum computer for $U=4$ to demonstrate that it can also obtain the self-consistent value of $V$ to a good accuracy. We use the quantum chemistry package EUMEN by Cambridge Quantum Computing (CQC) in combination with the $\tket$ compiler for circuit optimization, which is freely available to researchers \cite{tket}. To partly compensate for noise in the quantum hardware we apply standard corrections to the outputs for state preparation and measurement (SPAM) errors \cite{Yung}.

We run the experiments on superconducting qubits on the IBM quantum computers, and use 4096(8192) shots for a measurement on 4(2) qubits.  We use the SPAM library in CQC's $\tket$ compiler for the 2 qubit measurements~\cite{cowtan_et_al:LIPIcs:2019:10397}, and the equivalent Qiskit's Ignis library for the 4 qubit measurements\cite{Qiskit}.

The ion trap quantum computer at the University of Maryland is based on a chain of individual $^{171}$Yb$^{+}$ ions confined in a Paul trap \cite{Debnath2016p63,Landsman2019}. The native operations of the system are single qubit rotations, or R gates, and two-qubit entangling interactions, or XX gates, which are created by coupling any pair of qubits via the motional modes in the trap \cite{Choi2014}. This experiment is performed on a chain of seven ions, of which five are used as qubits. For the 4(2) qubit runs we use 4000(5000) shots per measurement. SPAM errors on the output distribution are corrected via the inverse of an independently measured state-to-state error matrix. Importantly, this error characterization scales linearly with system size. In Appendix \ref{sec:appendixFullData} we compare 4-qubit results without and with SPAM correction for superconducting and trapped ion qubits.

\begin{table}[t]
\caption{
Experimental data obtained on the IBM superconducting qubit quantum computer: the results are analogous to those obtained by classical simulation in Tab. \ref{tab:enesSimul}. The ``Optimal $\theta$'' results have been measured using the angles optimized with the simulator, while for the data in the ``VQE'' columns the VQE optimization loop has been performed on the IBM device, and for the ``DMFT+VQE'' column the whole DMFT loop has been performed on a quantum computer (converged value for $V$ is $V=0.755$). The value of $\lambda$ given for ``DMFT+VQE'' is the one used in the DMFT self-consistency loop based on the regularization sum rules (Eq. (\ref{eq:lambda})). \label{tab:enesIBM} 
}
\begin{tabular}{c|c|c|ccc}
\hline
         &                           &  4 qubits  & \multicolumn{3}{c}{2 qubits}  \\
         & Exact            & Optimal $\theta$       & Optimal $\theta$  & VQE  & DMFT+VQE   \\
\hline                                                                                                                          
 $E_0$      & -1.795          & -1.500        &  -1.700      & -1.823  &    -1.809     \\
 $E_{3,0}$  & -1.247          & -1.111        &  -1.259      & -1.248  &    -1.245     \\
 $E_{3,2}$  &  1.247          &  1.025        &   1.253      &  1.248  &     1.244     \\
$\lambda$   &  0.262          &  0.113        &   0.210      &  0.275  &     0.271     \\
\hline
\end{tabular}
\end{table}
\begin{table}[t]
\caption{
Experimental data obtained on the ion trap quantum computer at the University of  Maryland: the results are analogous to those obtained by classical simulation in Tab. \ref{tab:enesSimul}. The ``Optimal $\theta$'' results have been measured using the angles optimized with the simulator.
\label{tab:enesUMD}
}
\begin{tabular}{c|c|c|cc}
\hline
         &            & 4 qubits  & 2 qubits     \\
         &      Exact           &  Optimal $\theta$    &    Optimal $\theta$  \\
\hline                                                           
 $E_0$     & -1.795     &  -1.691     &   -1.742   \\
 $E_{3,0}$ & -1.247     &  -1.178     &   -1.208   \\
 $E_{3,2}$ &  1.247     &   1.144     &    1.230   \\
$\lambda$  &  0.262     &   0.224     &    0.258   \\
\hline
\end{tabular}
\end{table}

We first perform a measurement for fixed circuit parameter values set equal to the ones optimized on the classical simulator with VQE (Tab. \ref{tab:enesSimul}), using both the 4 qubit PT approach and the 2 qubit CR approach.
The results for the IBM quantum computer are given in Tab. \ref{tab:enesIBM}, and the ones for the quantum computer at the University of  Maryland are presented in Tab.\ref{tab:enesUMD} (more detailed results are given in Appendix \ref{sec:appendixFullData}).
The 4 qubit experiments on the IBM device give energies that are within about 10-20\% of the exact values, and for the 2 qubit CR experiments we obtain energies within about 5\% of the exact ones. For $\lambda$ the deviation is larger, about 20\%(60\%) from the exact values when using the CR(PT) method. This larger error is expected, since the circuit to evaluate $\lambda$ is longer than the one for the energies. Importantly, for a DMFT calculation $\lambda$ is obtained from the exact sum rule (Eq. (\ref{eq:lambda})), so that the accuracy of the final result is entirely determined by the accuracy of the energies. On the ion trap device, the measurement outcomes are closer to the exact values. The results on 4(2) qubits for the energy are within about 6-8\%(1-3\%) of the exact ones, and $\lambda$ for the CR method essentially matches the exact value (Tab. \ref{tab:enesUMD}).
Note that here we consider the ph-symmetric case, since it allows to reduce the number of computations. To show that the approach is valid also for the general case, in Appendix \ref{sec:AppendixNPHS} we present an example for a simulation and experiment for the general non-particle-hole symmetric case.

These results show that the required energies can be calculated rather accurately with experiments on the considered quantum devices. They also confirm that the CR approach generally gives results that are more noise-resilient. We therefore apply the CR method to perform a full VQE loop on the IBM quantum computer. 
For this experiment we perform the VQE optimization with the Rotosolve optimization algorithm.\cite{rotosolve}
The VQE results in Tab. \ref{tab:enesIBM} show that it converges to within 2\% of the exact results. The value of $\lambda$ is accurate to about 5\%. These results are better than those obtained at the theoretical optimal angles on the same device, and are therefore indicative that the VQE optimization on the quantum device partly compensates the calibration errors in the hardware.

So far in order to reduce the runtime of the experiments we have only presented VQE results for a fixed $V$ at exact DMFT self-consistency. The experimental time for a full DMFT loop is the product of the single VQE loop time, and the number of iterations in the outer DMFT loop to update $V$ to self-consistency (Fig. \ref{fig:DMFTSchematic}). Given the high accuracy of the VQE results we expect that a full DMFT loop will exhibit similar convergence as the simulated one, and therefore reach self-consistency at about ten iterations (Fig. \ref{fig:DMFTLoop}a). We verify this by running a full DMFT loop with the 2-qubit CR method on the IBM quantum computer. The results are shown in Fig. \ref{fig:DMFTLoop}a, and indeed the convergence of the DMFT loop on the quantum hardware is very similar to the simulated results. Convergence to a set tolerance of $| V - \sqrt{z} | < 0.01$ is reached after 7 iterations, and the final self-consistent value for $V$ is $V=0.755$, and therefore very close to the exact one. In the last column of Tab. \ref{tab:enesIBM} we also show that the resulting energies and $\lambda$ are close to the exact values (in the DMFT loop $\lambda$ is calculated with the regularization Eq. (\ref{eq:lambda})).
Note that due to the lower accuracy of the 4-qubit PT approach on the IBM quantum computer we expect worse convergence of the DMFT loop on 4 qubits when compared to the 2-qubit CR approach, while on the trapped ion quantum computer a similar DMFT convergence can be expected also on 4 qubits due to the higher accuracy.

\begin{figure}[t]
\centering\includegraphics[width=0.47\textwidth,clip=true]{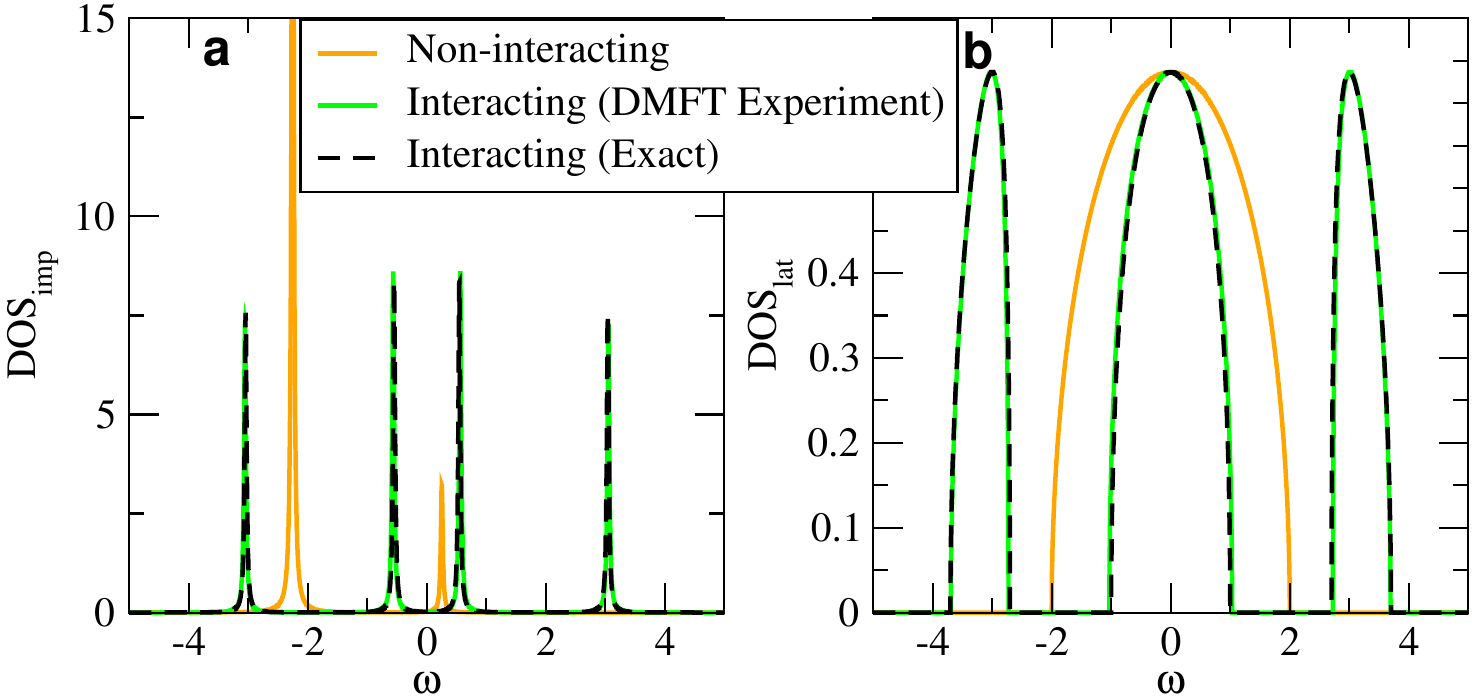}
\caption{(a) DOS on the impurity site and (b) corresponding DMFT DOS on the Bethe lattice for $U=4$, computed with many-electron interactions as a DMFT+VQE experiment using the 2 qubit reduced circuit approach on the IBM quantum computer (green curve). We use the output of the DMFT self-consistent loop on the quantum computer for $V$ ($V=0.755$) The results of the experiment compare well with the exact values at the exact $V=0.745356$ (dashed curve). For comparison we also present the DOS without many-electron interactions (orange curve).
}
\label{fig:DOS}
\end{figure}
Having ran the DMFT self-consistency on the quantum computer to obtain $V$, one can then determine the electronic structure of the system. Using the energies calculated with VQE, and $\lambda$ regularized using Eq. (\ref{eq:lambda}), we plot the DOS on the impurity site and on the corresponding Bethe lattice (Fig. \ref{fig:DOS}). The agreement with the exact results is good, the small deviations are caused by the small differences in the energies of the VQE experiment and the exact values. For comparison we also present the DOS obtained for the non-interacting system, which for the Bethe lattice has only a single broad peak, and therefore lacks the correct three peak structure obtained with DMFT.

\section{Conclusions}
We have presented an algorithm that performs DMFT calculations on currently available quantum computers. Our benchmarks on superconducting and trapped ion qubits for 2-site DMFT show that such practical calculations are possible with low levels of error. The reason is that the method is based on VQE total energy calculations, which are generally more resilient to noise than the real time evolution of states. The method will extend to growing system sizes compatible with near term quantum devices, and will benefit from the development for VQE calculations for quantum chemistry calculations of molecules, since the required quantum circuits are similar. We expect that our proof of concept demonstration that DMFT can be run on current quantum hardware will spark additional research into quantum algorithms for condensed matter physics systems.

\section{Acknowledgments}
IR acknowledges financial support from the UK Department of Business, Energy and Industrial Strategy (BEIS). IR/RD/DR/NF acknowledge financial support from Innovate UK through the Analysis for Innovators Scheme A4I R3 Mini Project 104936. HC acknowledges the support though a Teaching Fellowship from UCL, LW acknowledges kindly the support through the Google PhD Fellowship in Quantum Computing.
CHA acknowledges financial support from CONACYT (doctoral grant no. 455378). NML acknowledges financial support from the NSF Physics Frontier Center at JQI (grant no. PHY-1430094). EG is supported by the UK EPSRC [EP/P510270/1]. AP is supported by the InQUBATE Training and Skills Hub grant EPSRC EP/P510270/1. AP and IR thank the members of the EPSRC (grant No. EP/S005021/1) Prosperity Partnership in Quantum Software for Modeling and Simulation for useful discussions.

\appendix

\section{Matrix elements relations}
\label{sec:AppendixLambda}
In Sec. \ref{sec:dmft} we have used a standard Jordan-Wigner transform and as part of it we have introduced a modified form of the Pauli ladder operators that takes into account the fermionic nature of the electrons as
\begin{align}
\hat{\sigma}_{\alpha}^\pm&=\left(\prod_{\beta=1}^{\alpha-1}\hat{\sigma}_\beta^z\right) \frac{1}{2} \left(\hat{\sigma}_\alpha^x \pm i \hat{\sigma}_\alpha^y\right).
\label{eq:appendixSigmapm}
\end{align}
With this definition $\hat{\sigma}_\alpha^-$ ($\hat{\sigma}_\alpha^+$) creates (destroys) an electron on spin orbital $\alpha$. The matrix elements $\lambda_{\mathrm{p/h},\alpha,n}$ are then generally given by
\begin{align}
\lambda_{\mathrm{h},\alpha,n}&=\left|\left<\psi_{N_0-1,n}\left|\hat{\sigma}_\alpha^+\right|\psi_0\right>\right|^2,\label{eq:Appendixalphah}\\
\lambda_{\mathrm{p},\alpha,n}&=\left|\left<\psi_{N_0+1,n}\left|\hat{\sigma}_\alpha^-\right|\psi_0\right>\right|^2.\label{eq:Appendixalphap}
\end{align}
Using the definition in Eq. (\ref{eq:appendixSigmapm}) we obtain
\begin{align}
\left(\prod_{\beta=1}^{\alpha-1}\hat{\sigma}_\beta^z\right)\hat{\sigma}_\alpha^x&=
\hat{\sigma}_\alpha^+ +\hat{\sigma}_\alpha^-,\label{eq:Appendixr1}\\
i \left(\prod_{\beta=1}^{\alpha-1}\hat{\sigma}_\beta^z\right)\hat{\sigma}_\alpha^y&=
\hat{\sigma}_\alpha^+ -\hat{\sigma}_\alpha^-.
\end{align}
Furthermore, since $\hat{\sigma}_\alpha^-$ ($\hat{\sigma}_\alpha^+$) creates (destroys) one electron on spin orbital $\alpha$, we also have
\begin{align}
\left<\psi_{N',m}\left|\hat{\sigma}_\alpha^+\right|\psi_{N,n}\right> &= \delta_{N',N-1}\left<\psi_{N-1,m}\left|\hat{\sigma}_\alpha^+\right|\psi_{N,n}\right>,\\
\left<\psi_{N',m}\left|\hat{\sigma}_\alpha^-\right|\psi_{N,n}\right> &= \delta_{N',N+1}\left<\psi_{N+1,m}\left|\hat{\sigma}_\alpha^-\right|\psi_{N,n}\right>\label{eq:Appendixme2}.
\end{align}
Combining Eqs. (\ref{eq:Appendixr1}-\ref{eq:Appendixme2}) we obtain
\begin{align}
\left<\psi_{N-1,m}\left|\hat{\sigma}_\alpha^+\right|\psi_{N,n}\right>=
\left<\psi_{N-1,m}\left|\left(\prod_{\beta=1}^{\alpha-1}\hat{\sigma}_\beta^z\right)\hat{\sigma}_\alpha^x\right|\psi_{N,n}\right>,\label{eq:AppendixSp}\\
\left<\psi_{N+1,m}\left|\hat{\sigma}_\alpha^-\right|\psi_{N,n}\right>=
\left<\psi_{N+1,m}\left|\left(\prod_{\beta=1}^{\alpha-1}\hat{\sigma}_\beta^z\right)\hat{\sigma}_\alpha^x\right|\psi_{N,n}\right>,\label{eq:AppendixSm}
\end{align}
and equivalently
\begin{align}
\left<\psi_{N-1,m}\left|\hat{\sigma}_\alpha^+\right|\psi_{N,n}\right>=
i\left<\psi_{N-1,m}\left|\left(\prod_{\beta=1}^{\alpha-1}\hat{\sigma}_\beta^z\right)\hat{\sigma}_\alpha^y\right|\psi_{N,n}\right>,\\
\left<\psi_{N+1,m}\left|\hat{\sigma}_\alpha^-\right|\psi_{N,n}\right>=
-i\left<\psi_{N+1,m}\left|\left(\prod_{\beta=1}^{\alpha-1}\hat{\sigma}_\beta^z\right)\hat{\sigma}_\alpha^y\right|\psi_{N,n}\right>.
\end{align}
By inserting Eqs. (\ref{eq:AppendixSp}-\ref{eq:AppendixSm}) into Eqs. (\ref{eq:Appendixalphah}-\ref{eq:Appendixalphap}) we obtain the relations for the matrix elements  $\lambda_{\mathrm{p/h},\alpha,n}$ given in Eqs. (\ref{eq:alphah}) and Eq. (\ref{eq:alphap}).

\section{Circuit reduction}
\label{sec:appendixCR}

\subsection{General relations}
Here we perform the transformations to reduce the number of qubits within 2-site DMFT in order to restrict the wavefunction to a fixed number of electrons and total $z$ component of the spin. Such a restriction is possible, since the Hamiltonian in Eq. (\ref{eq:H2s}) commutes with both the total number operator
\begin{equation}
\hat{N}=\hat{n}_1 + \hat{n}_2 + \hat{n}_3 + \hat{n}_4,
\end{equation}
and the operator for the total $z$ component of the spin (in units of $\hbar/2$)
\begin{equation}
\hat{S}_z=\hat{n}_1 + \hat{n}_2 - \hat{n}_3 - \hat{n}_4,
\end{equation}
where $\hat{n}_{\alpha}=\hat{\sigma}_{\alpha}^- \hat{\sigma}_{\alpha}^+$ is the number operator on spin orbital $\alpha$. As outlined in the main text, qubits 1 and 2 correspond to $\uparrow$ spin orbitals, while qubits 3 and 4 correspond to $\downarrow$ spin orbitals. Note that in this section we use Greek subscripts for both the bath and interacting sites. As outlined in Sec. \ref{sec:2sDMFTModel}, the qubit indices are mapped to site and spin indices as follows: qubit 1 (2) represents $\uparrow$ electrons on site 1 (2), while qubit 3 (4) represents a $\downarrow$ electron on site 1 (2). 

The Pauli operators satisfy the commutator relation
\begin{equation}
\left[\hat{\sigma}^a_\alpha,\hat{\sigma}^b_\beta\right]=2 i \delta_{\alpha\beta} \epsilon_{abc} \hat{\sigma}^c_\alpha, 
\label{eq:commPauli}
\end{equation}
where $\epsilon_{abc}$ is the Levi-Civita symbol, and the anti-commutator relation
\begin{equation}
\lbrace\hat{\sigma}^a_\alpha,\hat{\sigma}^b_\alpha\rbrace=2 \delta_{ab}. 
\label{eq:anticommPauli}
\end{equation}
Using these relations one obtains
\begin{equation}
\hat{\sigma}^a_\alpha \hat{\sigma}^b_\alpha=\delta_{ab} +i \epsilon_{abc} \hat{\sigma}^c_\alpha. 
\end{equation}
With this equation we simplify the number operator for a spin orbital to
\begin{align}
\hat{n}_{\alpha}&=\hat{\sigma}_{\alpha}^- \hat{\sigma}_{\alpha}^+= \frac{1}{4}\left(\hat{\sigma}_{\alpha}^x-i\hat{\sigma}_{\alpha}^y\right) \left(\hat{\sigma}_{\alpha}^x+i\hat{\sigma}_{\alpha}^y\right)\nonumber\\
&=\frac{1}{2}\left(1-\hat{\sigma}^z_{\alpha}\right).
\end{align}
We can therefore rewrite $\hat{N}$ and $\hat{S}_z$ as
\begin{align}
\hat{N}&=2-\frac{1}{2}\left(\hat{\sigma}^z_1 +\hat{\sigma}^z_2 +\hat{\sigma}^z_3 +\hat{\sigma}^z_4\right),
\label{eq:nop}\\
\hat{S}_z&=\frac{1}{2}\left(-\hat{\sigma}^z_1 -\hat{\sigma}^z_2 +\hat{\sigma}^z_3 +\hat{\sigma}^z_4\right).
\label{eq:sop}
\end{align}

One can then verify that the commutators between $\hat{H}$ (Eq. (\ref{eq:H2s})), $\hat{N}$, and $\hat{S}_z$ vanish
\begin{equation}
\left[\hat{H},\hat{N}\right]=\left[\hat{H},\hat{S}_z\right]=\left[\hat{N},\hat{S}_z\right]=0.
\end{equation}
We can therefore evaluate the eigenstates of the Hamiltonian separately for each given number of electrons, $N$, and total $z$ component of the spin, $S_z$. In the following subsections we derive effective Hamiltonians with a reduced number of qubits for each pair of $N$ and $S_z$. As a matter of notation, we denote the expectation value of an operator by removing the hat from the symbol of the operator, so that for example $\sigma_1^z=\langle\hat{\sigma}_1^z\rangle$. We then denote a basis vector of the system by the list of $\sigma_\alpha^z$ values within the ket: $\left.|\sigma^z_1,\sigma^z_2,\sigma^z_3,\sigma^z_4\right>$.

\subsection{\textit{N}=0}
\label{sec:appendixCR0e}

For $N=0$ from Eq. (\ref{eq:nop}) we obtain the condition $\sigma^z_1+\sigma^z_2+\sigma^z_3 + \sigma^z_4 = 4$. Since $|\sigma^z_\alpha|=1$ this condition implies that $\sigma^z_1=\sigma^z_2=\sigma^z_3=\sigma^z_4=1$, resulting also to $S_z=0$. Therefore we have only one possible state, with energy $E_{N=0,S_z=0}=\left<1,1,1,1|\right.\hat{H}\left.|1,1,1,1\right>=\mu-U/4-\epsilon_2$. 

\subsection{\textit{N}=1}
\label{sec:appendixCR1e}
For $N=1$ Eq. (\ref{eq:nop}) gives the condition $\sigma^z_3 + \sigma^z_4 = 2 -\sigma^z_1-\sigma^z_2$. With Eq. (\ref{eq:sop}) we then have $S_z=1-\sigma^z_1-\sigma^z_2$. Since in general $\sigma^z_3 + \sigma^z_4 \le 2$ we also have the additional relation $0\le \sigma^z_1+\sigma^z_2\le 2$. Since $|\sigma^z_\alpha|=1$, with these conditions the possible values for $S_z$ are $\lbrace-1,1\rbrace$. This is generally the case for an arbitrary number of qubits, where one adds or removes a single electron from a $S_z=0$ state. The operator $\hat{S}_z$ therefore determines the spin-imbalance of such systems, and to fully characterize the state one only needs to add the information about the electron distribution across the various sites. We therefore introduce the new operators to uniquely determine the state:
\begin{align}
\hat{\bar{\sigma}}^z_1=\frac{1}{2}\left(\hat{\sigma}^z_1 +\hat{\sigma}^z_2 -\hat{\sigma}^z_3 -\hat{\sigma}^z_4\right),\label{eq:1esigmaz}\\
\hat{\bar{\sigma}}^z_2=\frac{1}{2}\left(\hat{\sigma}^z_1 -\hat{\sigma}^z_2 +\hat{\sigma}^z_3 -\hat{\sigma}^z_4\right),\label{eq:2esigmaz}
\end{align}
where $\hat{\bar{\sigma}}^z_1=-\hat{S}_z$ determines the spin-imbalance, and $\hat{\bar{\sigma}}^z_2$ determines the imbalance of electron number across the two sites. Raising or lowering the expectation values of these states involves spin-flips for the first operator, and spin-preserving electron hops between sites for the second operator. This corresponds to the ladder operators
\begin{align}
\hat{\bar{\sigma}}_1^+&=\hat{\sigma}_3^-\hat{\sigma}_1^++\hat{\sigma}_4^-\hat{\sigma}_2^+,\\
\hat{\bar{\sigma}}_2^+&=\hat{\sigma}_2^-\hat{\sigma}_1^++\hat{\sigma}_4^-\hat{\sigma}_3^+,
\end{align}
and resulting $x$ and $y$ operators
\begin{align}
\hat{\bar{\sigma}}_1^x&=\frac{1}{2}\left(\hat{\sigma}_1^x\hat{\sigma}_3^x+\hat{\sigma}_1^y\hat{\sigma}_3^y+\hat{\sigma}_2^x\hat{\sigma}_4^x+\hat{\sigma}_2^y\hat{\sigma}_4^y\right),\label{eq:1esigma1x}\\
\hat{\bar{\sigma}}_2^x&=\frac{1}{2}\left(\hat{\sigma}_1^x\hat{\sigma}_2^x+\hat{\sigma}_1^y\hat{\sigma}_2^y+\hat{\sigma}_3^x\hat{\sigma}_4^x+\hat{\sigma}_3^y\hat{\sigma}_4^y\right),\\
\hat{\bar{\sigma}}_1^y&=\frac{1}{2}\left(-\hat{\sigma}_1^x\hat{\sigma}_3^y+\hat{\sigma}_1^y\hat{\sigma}_3^x-\hat{\sigma}_2^x\hat{\sigma}_4^y+\hat{\sigma}_2^y\hat{\sigma}_4^x\right),\\
\hat{\bar{\sigma}}_2^y&=\frac{1}{2}\left(-\hat{\sigma}_1^x\hat{\sigma}_2^y+\hat{\sigma}_1^y\hat{\sigma}_2^x-\hat{\sigma}_3^x\hat{\sigma}_4^y+\hat{\sigma}_3^y\hat{\sigma}_4^x\right).\label{eq:1esigma2y}
\end{align}
It can be verified that for all states with $N=1$ these new operators satisfy the commutator relations Eqs. (\ref{eq:commPauli}-\ref{eq:anticommPauli}), and therefore can be represented as system of two qubits.

For a system with $N=1$, in order to have the single electron on site 1 with spin up we need to have $\bar{\sigma}_1^z=-S_z=-1$ and $\bar{\sigma}_2^z=-1$, while for all other eigenstates of the 2-qubit system there is no electron on site one with spin up. This allows us to construct the inverse projection as
\begin{align}
\hat{\sigma}_1^z&=-\frac{1}{2}\left(1-\hat{\bar{\sigma}}_1^z\right)\left(1-\hat{\bar{\sigma}}_2^z\right)+1.
\label{eq:1eIT1}
\end{align}
For the other sites and spins the inverse transformation can be constructed in an analogous way:
\begin{align}
\hat{\sigma}_2^z&=-\frac{1}{2}\left(1-\hat{\bar{\sigma}}_1^z\right)\left(1+\hat{\bar{\sigma}}_2^z\right)+1,\\
\hat{\sigma}_3^z&=-\frac{1}{2}\left(1+\hat{\bar{\sigma}}_1^z\right)\left(1-\hat{\bar{\sigma}}_2^z\right)+1,\\
\hat{\sigma}_4^z&=-\frac{1}{2}\left(1+\hat{\bar{\sigma}}_1^z\right)\left(1+\hat{\bar{\sigma}}_2^z\right)+1.
\label{eq:1eIT4}
\end{align}
It is straight forward to verify that these relations do indeed give Eqs. (\ref{eq:1esigmaz}-\ref{eq:2esigmaz}) when solved for $\hat{\bar{\sigma}}_1^z$ and $\hat{\bar{\sigma}}_2^z$.

With Eqs. (\ref{eq:1esigma1x}-\ref{eq:1eIT4}) we can transform the 4-qubit Hamiltonian (Eq. \ref{eq:H2s}) for $N=1$ to the 2-qubit system as
\begin{align}
\hat{H}=\left(\frac{\mu}{2}+\frac{\epsilon_2}{2}\right)\hat{\bar{\sigma}}_2^z+V \hat{\bar{\sigma}}_2^x+\left(\frac{\mu}{2}-\frac{U}{4}-\frac{\epsilon_2}{2}\right).
\label{eq:H22Q1e}
\end{align}

\subsection{\textit{N}=2}
\label{sec:appendixCR2e}

For $N=2$ from Eq. (\ref{eq:nop}) we obtain the condition $\sigma^z_3 + \sigma^z_4 = -\sigma^z_1-\sigma^z_2$, so that with Eq. (\ref{eq:sop}) $S_z=-\sigma^z_1-\sigma^z_2$. Since $|\sigma^z_\alpha|=1$, the possible values for $S_z$ are $\lbrace-2,0,2\rbrace$. For $S_z=2$ there is only one possible state with $\sigma^z_1=\sigma^z_2=-1$, with energy $E_{N=2,S_z=2}=\left<-1,-1,1,1|\right.\hat{H}\left.|-1,-1,1,1\right>=-U/4$. Analogously, for $S_z=-2$ we only have the state with $\sigma^z_1=\sigma^z_2=1$, with energy $E_{N=2,S_z=-2}=-U/4$.

For $S_z=0$ we have the additional condition $\sigma^z_2=-\sigma^z_1$, which since $N=2$ also implies $\sigma^z_4=-\sigma^z_3$. We introduce a new set of Pauli operators, which guarantee that these conditions are automatically satisfied:
\begin{align}
\hat{\bar{\sigma}}_1^z&=\hat{\sigma}_1^z,\\
\hat{\bar{\sigma}}_2^z&=\hat{\sigma}_3^z,\\
\hat{\bar{\sigma}}_1^x&=\hat{\sigma}_1^x\hat{\sigma}_2^x,\\
\hat{\bar{\sigma}}_2^x&=\hat{\sigma}_3^x\hat{\sigma}_4^x,\\
\hat{\bar{\sigma}}_1^y&=\hat{\sigma}_1^y\hat{\sigma}_2^x,\\
\hat{\bar{\sigma}}_2^y&=\hat{\sigma}_3^y\hat{\sigma}_4^x.
\end{align}
The $z$ components are equal to the ones of the original $z$-components for qubit 1 and 3, while the $x$ and $y$ components also make sure that when the values of qubits 1 and 3 are flipped, the values of qubits 2 and 4 are flipped at the same time to satisfy the conditions above. One can verify that the new operators satisfy the commutator relations for Pauli operators (Eqs. (\ref{eq:commPauli}-\ref{eq:anticommPauli})), which shows that they can be represented as new system of two qubits. Within this subspace with $N=2$ and $S_z=0$ the inverse operations can be directly obtained from the equations above, with the additional relations for qubits 2 and 4 given by
\begin{align}
\hat{\sigma}_2^z&=-\hat{\bar{\sigma}}_1^z,\\
\hat{\sigma}_4^z&=-\hat{\bar{\sigma}}_2^z.
\end{align}
With these relations, and the commutator relations for Pauli operators (Eqs. (\ref{eq:commPauli}-\ref{eq:anticommPauli})), we can represent the Hamiltonian from Eq. (\ref{eq:H2s}) with the newly introduced transformed Pauli operators as
\begin{align}
\hat{H}&=\frac{U}{4}\hat{\bar{\sigma}}_1^z \hat{\bar{\sigma}}_2^z+\left(\frac{\mu}{2}-\frac{U}{4}+\frac{\epsilon_2}{2}\right)\left(\hat{\bar{\sigma}}_1^z+ \hat{\bar{\sigma}}_2^z\right)\nonumber\\
&+V\left(\hat{\bar{\sigma}}_1^x +\hat{\bar{\sigma}}_2^x\right).
\label{eq:H2s2e}
\end{align}
We can therefore evaluate the states for $N=2$ and $S_z=0$ with this reduced Hamiltonian on 2 qubits.

\subsection{\textit{N}=3}
\label{sec:appendixCR3e}
For $N=3$ Eq. (\ref{eq:nop}) gives the condition $\sigma^z_3 + \sigma^z_4 = -2 -\sigma^z_1-\sigma^z_2$. With Eq. (\ref{eq:sop}) we then have $S_z=-1-\sigma^z_1-\sigma^z_2$. Since in general $-2 \le \sigma^z_3 + \sigma^z_4$, this results to the additional relation $0\ge \sigma^z_1+\sigma^z_2\ge -2$. Since $|\sigma^z_\alpha|=1$, with these conditions the possible values for $S_z$ are $\lbrace-1,1\rbrace$.
For $N=3$ we can therefore use the same mapping to 2-qubits as for the 1-electron system, given in Eqs. (\ref{eq:1esigmaz}-\ref{eq:2esigmaz}) and Eqs. (\ref{eq:1esigma1x}-\ref{eq:1esigma2y}). The inverse mapping however needs to be modified to the appropriate states with $N=3$.

For a system with $N=3$, in order to have no electron on site one with spin up we need to have $\bar{\sigma}_1^z=-S_z=1$ and $\bar{\sigma}_2^z=1$, while for all other eigenstates of the 2-qubit system there is an electron on site one with spin up. We can impose this condition by constructing the inverse projection for this site and spin as
\begin{align}
\hat{\sigma}_1^z&=\frac{1}{2}\left(1+\hat{\bar{\sigma}}_1^z\right)\left(1+\hat{\bar{\sigma}}_2^z\right)-1.\label{eq:3eIT1}
\end{align}
For the other sites and spins the inverse transformation can be constructed in an analogous way:
\begin{align}
\hat{\sigma}_2^z&=\frac{1}{2}\left(1+\hat{\bar{\sigma}}_1^z\right)\left(1-\hat{\bar{\sigma}}_2^z\right)-1,\\
\hat{\sigma}_3^z&=\frac{1}{2}\left(1-\hat{\bar{\sigma}}_1^z\right)\left(1+\hat{\bar{\sigma}}_2^z\right)-1,\\
\hat{\sigma}_4^z&=\frac{1}{2}\left(1-\hat{\bar{\sigma}}_1^z\right)\left(1-\hat{\bar{\sigma}}_2^z\right)-1.\label{eq:3eIT4}
\end{align}
It can be verified that these relations give Eqs. (\ref{eq:1esigmaz}-\ref{eq:2esigmaz}) when solved for $\hat{\bar{\sigma}}_1^z$ and $\hat{\bar{\sigma}}_2^z$.

With Eqs. (\ref{eq:1eIT1}-\ref{eq:3eIT4}) we can transform the 4-qubit Hamiltonian (Eq. \ref{eq:H2s}) for $N=3$ to the 2-qubit system as
\begin{align}
\hat{H}=\left(\frac{\mu}{2}+\frac{\epsilon_2}{2}-\frac{U}{2}\right)\hat{\bar{\sigma}}_2^z+V \hat{\bar{\sigma}}_2^x-\left(\frac{\mu}{2}-\frac{U}{4}-\frac{\epsilon_2}{2}\right).
\label{eq:H22Q3e}
\end{align}

\subsection{\textit{N}=4}
\label{sec:appendixCR4e}

For $N=4$ from Eq. (\ref{eq:nop}) we obtain the condition $\sigma^z_1+\sigma^z_2+\sigma^z_3 + \sigma^z_4 = -4$. Since $|\sigma^z_\alpha|=1$ this condition implies that $\sigma^z_1=\sigma^z_2=\sigma^z_3=\sigma^z_4=-1$, resulting also to $S_z=0$. Therefore we have only one possible state, with energy $E_{N=4,S_z=0}=\left<-1,-1,-1,-1|\right.\hat{H}\left.|-1,-1,-1,-1\right>=-\mu+3U/4+\epsilon_2$.

\subsection{Matrix elements with circuit reduction}
\label{sec:appendixCRlambda}

On 4 qubits the hole matrix elements $\lambda_{\mathrm{h},1,n}$ of the Green's function (Eq. (\ref{eq:g})) are given by
\begin{equation}
\lambda_{\mathrm{h},1,n}=\left|\left<0\left|\hat{U}^\dagger_{2,0}\hat{\sigma}_1^-\hat{U}_{1,n}\right|0\right>\right|^2.
\end{equation}
This would generally require a back-mapping from the 2-qubit states to 4-qubits. However, if the ordering of the states in the reduced circuits is chosen appropriately, then one can perform this operation entirely on 2 qubits. This is indeed the case for our projections for 2 and 1 electrons, where the $\hat{\bar{\sigma}}^-$ operator applied to the 1-electron states, and when using the 2-electron inverse mapping for $S_z=0$, gives the correct 2-electron states. We can therefore evaluate $\lambda_{\mathrm{h},1,n}$ as
\begin{equation}
\lambda_{\mathrm{h},1,n}=\left|\left<\bar{0}\left|\hat{\bar{U}}^\dagger_{2,0}\hat{\bar{\sigma}}_1^-\hat{\bar{U}}_{1,n}\right|\bar{0}\right>\right|^2,
\end{equation}
where the bars over the quantities indicate that these are evaluated on 2 qubits. Since the 2-electron ground state has $S_z=0$, for $\lambda_{\mathrm{h},1,n}$ to be non-zero the 1-electron state needs to have $S_z=-1$, which implies $\bar{\sigma}_1^z=1$ for the 1-electron state. The 1-electron states $\hat{\bar{U}}_{1,n}\left.|\bar{0}\right>$ with $\bar{\sigma}_1^z=-1$ therefore have $\lambda_{\mathrm{h},1,n}=0$. For the 1-electron states with $S_z=-1$ we have $\hat{\bar{\sigma}}_1^-\hat{\bar{U}}_{1,n}\left.|\bar{0}\right>=\hat{\bar{\sigma}}_1^x\hat{\bar{U}}_{1,n}\left.|\bar{0}\right>$, so that
\begin{equation}
\lambda_{\mathrm{h},1,n}=\left|\left<\bar{0}\left|\hat{\bar{U}}^\dagger_{2,0}\hat{\bar{\sigma}}_1^x\hat{\bar{U}}_{1,n}\right|\bar{0}\right>\right|^2.
\end{equation}
Note that with the circuit of Fig. 2c one always has $\bar{\sigma}_1^z=1$, and therefore the states with finite $\lambda_{\mathrm{h},1,n}$. Therefore $\lambda_{\mathrm{h},1,n}$ can be evaluated entirely on the 2-qubit circuit with this equation.

The particle matrix elements $\lambda_{\mathrm{p},1,n}$ of the Green's function on 2 qubits can be obtained in an analogous way, and are given by
\begin{equation}
\lambda_{\mathrm{p},1,n}=\left|\left<\bar{0}\left|\hat{\bar{U}}^\dagger_{2,0}\hat{\bar{\sigma}}_1^+\hat{\bar{U}}_{3,n}\right|\bar{0}\right>\right|^2.
\end{equation}
Since the 2-electron ground state has $S_z=0$, for $\lambda$ to be non-zero the 3-electron state needs to have $S_z=1$, which implies $\bar{\sigma}_1^z=-1$ for the 3-electron state. The 3-electron states $\hat{\bar{U}}_{3,n}\left.|\bar{0}\right>$ with $\bar{\sigma}_1^z=1$ therefore have $\lambda_{\mathrm{p},1,n}=0$. For the 3-electron states with $S_z=1$ we have $\hat{\bar{\sigma}}_1^+\hat{\bar{U}}_{3,n}\left.|\bar{0}\right>=\hat{\bar{\sigma}}_1^x\hat{\bar{U}}_{3,n}\left.|\bar{0}\right>$, so that
\begin{equation}
\lambda_{\mathrm{p},1,n}=\left|\left<\bar{0}\left|\hat{\bar{U}}^\dagger_{2,0}\hat{\bar{\sigma}}_1^x\hat{\bar{U}}_{3,n}\right|\bar{0}\right>\right|^2.
\end{equation}
Note that with the circuit of Fig. 2c one always has $\bar{\sigma}_1^z=1$, and therefore the states with $\lambda_{\mathrm{p},1,n}=0$. To obtain the state at the same energy with $\sigma_1^z=-1$, which has finite $\lambda_{\mathrm{h},1,n}$, one simply needs to add a $R_y$ rotation by $\pi$ to the first qubit. In this way we can perform the entire calculation of the $\lambda_{\mathrm{p},1,n}$ on the 2-qubit system.

\section{Additional 4-qubit experimental data}
\label{sec:appendixFullData}

As outlined in Sec. \ref{sec:qci} the particle-hole symmetric 2-site DMFT system is fully characterized by $E_{0}$, $E_{3,0}$, $E_{3,2}$, and $\lambda=\lambda_{p,1,0}$, and the results for these quantities measured on the quantum computers are given in Tabs. \ref{tab:enesIBM} and \ref{tab:enesUMD}. The remaining energies and matrix elements are then obtained by the exact relations $E_{3,1}=E_{3,0}$, $E_{3,3}=E_{3,2}$, $\lambda_{p,1,2}=1/2-\lambda$, and $\lambda_{p,1,1}=\lambda_{p,1,2}=0$. To evaluate to what extent these exact relations hold when measured on quantum computers, in Tab. \ref{tab:enesIBMUMDFull} we present the values of all these quantities measured on the quantum computers within the 4-qubit PT approach.
It can be seen that all relations are fulfilled to a good approximation.
Furthermore we also present the results without and with SPAM correction, to evaluate the effect of SPAM correction on the data. Overall it can be seen that SPAM correction tends to improve the data, although it is not always systematic, in particular the matrix elements $\lambda_{3,1,n}$ are not improved by adding SPAM corrections.  
\begin{table}[t]
\caption{
Experimental data obtained on the IBM superconducting qubit quantum computer (IBM QC) and on the trapped ion quantum computer at the University of Maryland (UMD QC), extending the 4-qubit results presented in Tabs. \ref{tab:enesIBM} and \ref{tab:enesUMD} for ``Optimal $\theta$''. We evaluate $E_{3,n}$ and $\lambda_{\mathrm{p},1,n}$ for all the $n\in\lbrace0,1,2,3\rbrace$ on the quantum hardware. Furthermore, we compare the raw result as output by the quantum computer with the SPAM corrected results.
\label{tab:enesIBMUMDFull}
}
\begin{tabular}{c|c|cc|cc}
\hline
         &            & \multicolumn{2}{c|}{IBM QC}  & \multicolumn{2}{c}{UMD QC}     \\
         &      Exact           &  \thead{Optimal $\theta$ \\ (raw result)}            &    \thead{Optimal $\theta$ \\ (SPAM-cor.)}      &    \thead{Optimal $\theta$ \\ (raw result)}        &    \thead{Optimal $\theta$ \\ (SPAM-cor.)}     \\
\hline                                                           
 $E_0$            &  -1.795      & -1.196  &  -1.500    & -1.603  & -1.691   \\
 $E_{3,0}$        &  -1.247      & -0.810  &  -1.111    & -1.111  & -1.178   \\
 $E_{3,1}$        &  -1.247      & -0.799  &  -0.981    & -1.055  & -1.112   \\
 $E_{3,2}$        &   1.247      &  1.001  &   1.025    &  1.129  &  1.144   \\
 $E_{3,3}$        &   1.247      &  0.934  &   0.972    &  1.048  &  1.098   \\
 $\lambda_{3,0}$  &   0.262      &  0.145  &   0.113    &  0.228  &  0.224   \\
 $\lambda_{3,1}$  &   0          &  0.010  &   0.002    &  0.002  &  0.002   \\
 $\lambda_{3,2}$  &   0.236      &  0.189  &   0.159    &  0.211  &  0.221   \\
 $\lambda_{3,3}$  &   0          &  0.011  &   0.004    &  0.001  &  0.000   \\
\hline
\end{tabular}
\end{table}

\section{Non-particle-hole symmetric system}
\label{sec:AppendixNPHS}
\begin{figure}[b]
\centering\includegraphics[width=0.47\textwidth,clip=true]{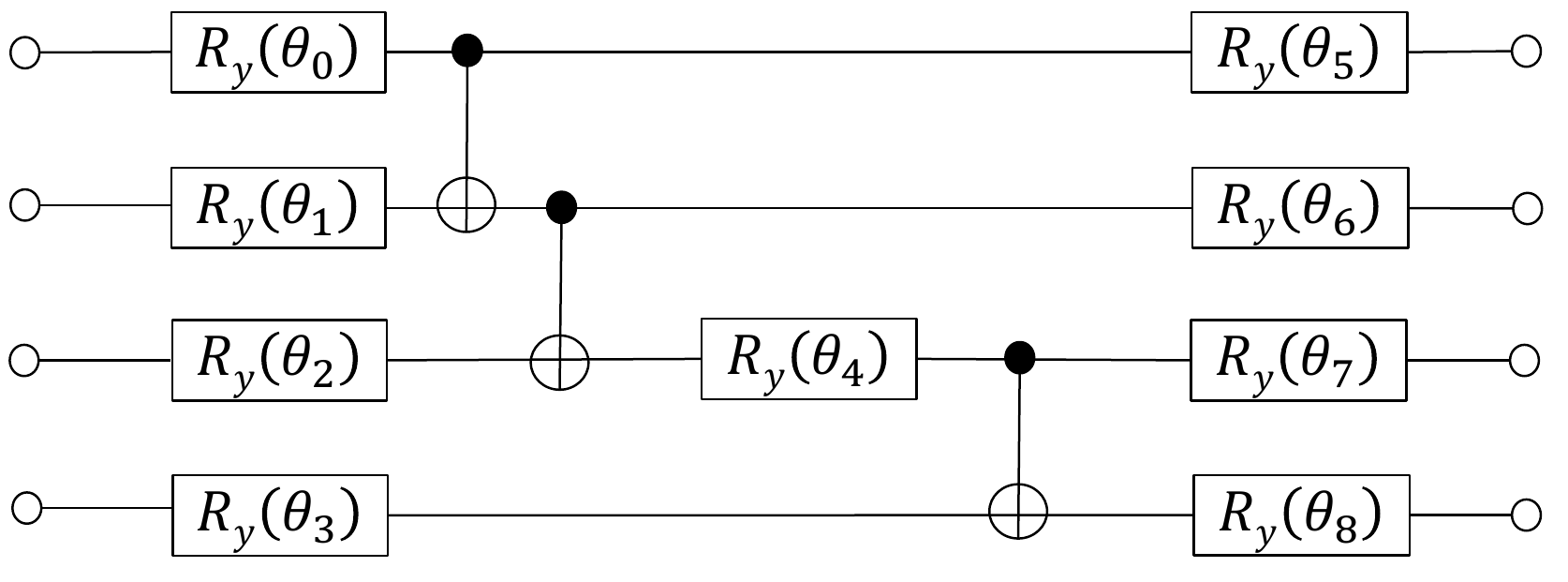}
\caption{Quantum circuit for the 4-qubit DMFT calculation based on the perturbed Hamiltonian (PT) approach used away from particle-hole symmetry. With this ansatz circuit we can reproduce the exact numerical results for all parameters used in the general 2-site DMFT Hamiltonian in Eq. (\ref{eq:H2s}).}
\label{fig:AppendixQC}
\end{figure}
The relations $E_{3,n}=E_{1,n}$ and $\lambda_{\mathrm{p},1,n}=\lambda_{\mathrm{h},1,n}$ are only valid at particle-hole (ph) symmetry. For a general set of parameters in the Hamiltonian of Eq. (\ref{eq:H2s}) this is not the case, and we need to compute the quantities for one and three electrons separately. These can be computed using the same methods as for the ph-symmetric case, so that moving away from ph-symmetry mainly leads to double the amount of computations that need to be performed on a quantum computer. Furthermore depending on the system it might be necessary to use a richer hardware efficient ansatz with more parameters in order to be able to describe all the states of the more general Hamiltonian. For our 2-qubit CR approach the circuits used for the ph-symmetric case (Figs. \ref{fig:QC}bc) also fully describe all states of the system away from ph-symmetry. For 4-qubits however the circuit shown in Fig. \ref{fig:QC}a can only approximately reproduce the exact numerical results. We therefore needed to extend the circuit, and we verified that the circuit shown in Fig. \ref{fig:AppendixQC} represents the smallest possible extension to the one of Fig. \ref{fig:QC} that can fully reproduce the exact results. Since the required extension is only one $R_y$ rotation, the expected quality of results is similar to that obtained for ph-symmetry. 
\begin{table}[h]
\caption{
Experimental data obtained on the ion trap quantum computer at the University of  Maryland, and compared to the numerically exact values. The ``Optimal $\theta$'' results have been measured using the angles for the circuit ansatz optimized with the simulator. For the 2-qubit results we compute the states for 1 and 3 electrons only for those cases that give a finite $\lambda_{\mathrm{p/h},1,n}$, the remaining states could be obtained by simply adding an additional $R_y(\pi)$ rotation on the first qubit.
\label{tab:enesUMDFullnonphs}
}
\begin{tabular}{c|c|cc|cc}
\hline
         &            & \multicolumn{2}{c|}{4 qubits}  & \multicolumn{2}{c}{2 qubits}     \\
         &      Exact           &  \thead{Optimal $\theta$ \\ (raw result)}            &    \thead{Optimal $\theta$ \\ (SPAM-cor.)}      &    \thead{Optimal $\theta$ \\ (raw result)}        &    \thead{Optimal $\theta$ \\ (SPAM-cor.)}     \\
\hline                                                           
 $E_0$            &  -1.837   &   -1.600   & -1.730    & -1.752  &  -1.792    \\
 $E_{1,0}$        &  -1.033   &   -0.843   & -0.927    & -1.010  &  -1.031    \\
 $E_{1,1}$        &  -1.033   &   -0.896   & -0.987    &        &           \\
 $E_{1,2}$        &   0.896   &    0.833   &  0.848    & 0.889  &  0.897    \\
 $E_{1,3}$        &   0.896   &    0.926   &  0.940    &        &           \\
 $\lambda_{1,0}$  &   0       &    0.003  &  0.001   &        &           \\
 $\lambda_{1,1}$  &   0.217   &    0.195   &  0.193   & 0.221  &  0.215    \\
 $\lambda_{1,2}$  &   0.033   &    0.044   &  0.041   & 0.038  &  0.035    \\
 $\lambda_{1,3}$  &   0       &    0.003  &  0.002   &        &           \\
 $E_{3,0}$        &  -0.624   &   -0.265   & -0.507    &-0.516   & -0.581    \\
 $E_{3,1}$        &  -0.624   &   -0.477   & -0.550    &        &           \\
 $E_{3,2}$        &   4.212   &    3.959   &  4.030    & 4.204   &  4.211     \\
 $E_{3,3}$        &   4.212   &    3.952   &  4.042    &        &           \\
 $\lambda_{3,0}$  &   0.644   &    0.584  &  0.589   & 0.621  &  0.621    \\
 $\lambda_{3,1}$  &   0       &    0.006  &  0.002   &        &           \\
 $\lambda_{3,2}$  &   0.106   &    0.118  &  0.113   & 0.109  &  0.101    \\
 $\lambda_{3,3}$  &   0       &    0.002  &  0.001   &        &           \\
\hline
\end{tabular}
\end{table}

As an example demonstration we use the case of $U=4$ and $\mu=-0.16016$. We perform the DMFT self-consistency on a simulator to satisfy the conditions in Eqs. (\ref{eq:DMFT1}-\ref{eq:DMFT2}), and find that this gives the self-consistent impurity problem parameters $\epsilon_2=-0.29764$ and $V=0.93709$. The resulting self-consistent DMFT impurity occupation is $n_\mathrm{imp}=0.5$ (at ph-symmetry we have $n_\mathrm{imp}=1$). In the same way as for the ph-symmetric case we use the ``Optimal $\theta$'' parameters computed on the simulator to set the parameters of the quantum circuit in the experiment, and evaluate the energies and matrix elements for these. We perform the experiments on the trapped ion quantum computer at the University of Maryland, using both the PT and CR approaches, and the results are given in Tab. \ref{tab:enesUMDFullnonphs}. The overall agreement between the experimental results and the exact data is analogous to the ph-symmetric case (Tab. \ref{tab:enesIBMUMDFull}), showing that the method can equally be applied on existing quantum hardware also away from ph-symmetry.


%

\end{document}